\documentclass[fleqn,usenatbib]{mnras}

\usepackage{newtxtext,newtxmath}
\usepackage[T1]{fontenc}
\DeclareRobustCommand{\VAN}[3]{#2}
\let\VANthebibliography\thebibliography
\def\thebibliography{\DeclareRobustCommand{\VAN}[3]{##3}\VANthebibliography}

\usepackage[normalem]{ulem}
\usepackage{graphicx, subfigure, algorithm, xcolor, algorithmic, amsmath, amssymb, float, rotating, enumerate}

\newcommand{\kmps}{\rm km~s\ensuremath{^{-1} }\,}
\newcommand{\Gaia}{{\it Gaia}\,}

\graphicspath{{./}{plots/}}

\title[Bar-spiral interaction]{
Gaia DR3 data consistent with a short bar connected to a spiral arm}

\author[Vislosky, Minchev, et al.]{
E.~Vislosky,$^{1,2}$\thanks{E-mail: evislosk@u.rochester.edu}
I.~Minchev,$^{1}$\thanks{E-mail: iminchev@aip.de}
S.~Khoperskov,$^{1}$
M. Martig,$^{3}$
T. Buck,$^{4,5}$
T. Hilmi,$^{6}$
B. Ratcliffe,$^{1}$ \newauthor
J. Bland-Hawthorn,$^{7,8}$
A.C. Quillen,$^{2}$
M. Steinmetz,$^{1}$
R. de Jong$^{1}$
\\
$^{1}$Leibniz-Institut f\"{u}r Astrophysik Potsdam (AIP), An der Sternwarte 16, D-14482, Potsdam, Germany\\
$^2$Department of Physics and Astronomy, University of Rochester, Rochester, NY 14627, USA\\
$^3$Astrophysics Research Institute, Liverpool John Moores University, 146 Brownlow Hill, Liverpool L3 5RF, UK\\
$^4$Universit\"at Heidelberg, Interdisziplin\"ares Zentrum f\"ur Wissenschaftliches Rechnen, Im Neuenheimer Feld 205, D-69120 Heidelberg, Germany\\
$^5$Universit\"at Heidelberg, Zentrum f\"ur Astronomie, Institut f\"ur Theoretische Astrophysik, Albert-Ueberle-Straße 2, D-69120 Heidelberg, Germany\\
$^6$Department of Physics, University of Surrey, Guildford GU2 7XH, UK\\
$^7$Sydney Institute for Astronomy, School of Physics, A28, The University of Sydney, NSW 2006, Australia\\
$^8$Centre of Excellence for All-Sky Astrophysics in Three Dimensions (ASTRO 3D), Australia
}

\date{Accepted 2023 December 22. Received 2023 December 22; in original form 2023 September 29}

\pubyear{2023}

\begin{document}
\label{firstpage}
\pagerange{\pageref{firstpage}--\pageref{lastpage}}
\maketitle

\begin{abstract}
We use numerical simulations to model \Gaia DR3 data with the aim of constraining the Milky Way bar and spiral structure parameters. We show that both the morphology and the velocity field in Milky Way-like galactic disc models are strong functions of time, changing dramatically over a few tens of Myr. This suggests that by finding a good match to the observed radial velocity field, $v_R(x,y)$, we can constrain the bar-spiral orientation. Incorporating uncertainties into our models is necessary to match the data; most importantly, a heliocentric distance uncertainty above 10-15\% distorts the bar's shape and $v_R$ quadrupole pattern morphology, and decreases its apparent angle with respect to the Sun-Galactocentric line. 
An excellent match to the \Gaia DR3 $v_R(x,y)$ field is found for a simulation with a bar length $R_b\approx3.6$ kpc. We argue that the data are consistent with a MW bar as short as $\sim3$~kpc, for moderate strength inner disc spiral structure ($A_2/A_0\approx0.25$) or, alternatively, with a bar length up to $\sim5.2$~kpc, provided that spiral arms are quite weak ($A_2/A_0\approx0.1$), and is most likely in the process of disconnecting from a spiral arm.
We demonstrate that the bar angle and distance uncertainty can similarly affect the match between our models and the data - a smaller bar angle (20$^\circ$ instead of 30$^\circ$) requires smaller distance uncertainty (20\% instead of 30\%) to explain the observations. {Fourier components of the face-on density distribution of our models suggest that the MW does not have strong m=1 and/or m=3 spirals near the solar radius.}
\end{abstract}


\section{Introduction}
\label{sec:intro}

The most significant stellar component of the Milky Way~(MW) disc is its central bar. Stellar bars are non-axisymmetric elongated features present in roughly two-thirds of disc galaxies in the local universe~\citep{knapen00, menendez07, marinova07, sheth08, masters11, erwin18}. However, the genuine properties of bars, such as their length, pattern speed, and strength, are difficult to resolve from the inner disc, bulge and spiral arms due to their mutual interconnection.

The MW bar was initially identified in near-infrared data \citep{blitz91} and in the study of gas kinematics \citep{binney91}. Due to the Sun's position within the Galactic plane, it has been difficult to study the bar directly from observations. At first, the MW bar was found to be quite short, with a half-length $R_b \sim 2.5$~kpc, by studying the peaks in the radial distribution of CO gas emission in the inner galaxy \citep{blitz91}, and by decomposing the stellar density distribution about the Galactic centre~(GC) \citep{weinberg92}. 

Refining indirect measurement techniques led to more consistent results, consistent with a fast, short bar with a pattern speed of $\Omega_b \sim 50-60$~$\rm km s^{-1} kpc^{-1}$ and $R_b \sim 3.5 $~kpc. These constraints were made by (1) matching hydrodynamic models of the ISM with Galactic $\rm H_I$ and molecular gas distributions in the inner MW using longitude-velocity ($\ell$-v) maps \citep{englmaier99, weiner99}, (2) matching the position of the Hercules moving group \citep{dehnen00, fux01, antoja12, monari17a} or the Pleiades and Sirius moving groups \citep{minchev10} in the \textit{Hipparcos} stellar velocity distribution, (3) reproducing the trend of the measured Oort's constant C with stellar velocity dispersion \citep{minchev07}, and (4) comparing to NIR stellar density distributions \citep{picaud03,lopez01}.

 More recently, direct observations of the inner MW disc have suggested, however, that the bar is actually longer and slower than previously expected ($R_b \sim 5$~ kpc and $\Omega_b \sim 35-45$~$\rm km s^{-1} kpc^{-1}$). This was determined by creating models for red clump magnitude distributions from NIR stellar surveys \citep{wegg15, portail17}, comparing MW $\ell$-v diagrams with hydrodynamical simulations including the effect of the bar \citep{sormani15, li16}, and explaining the Hercules moving group with the bar's corotation (CR) \citep{portail17, monari19} or the 4:1 Outer Lindblad Resonance (OLR) \citep{hunt18} of a long slow bar, rather than the 2:1 OLR in the case of a faster bar. 
 
 The transformative \Gaia DR2 and DR3 \citep{gaia18, drimmel23} datasets revealed arches, ridges, and streams in velocity and action space \citep{antoja18, kawata18, quillen18b, laporte19, bland-hawthorn19, brown21, queiroz21, poggio21}, showing unambiguously that the MW disc was out of equilibrium. This confirmed previous expectations based on incomplete pre-Gaia dataset (e.g., Hipparcos, RAVE, and SDSS), that a lot of disc phase-space structure could be explained as phase wrapping (mostly from the effect of the Sagittarius Dwarf galaxy, hereafter Sgr; \citealt{ibata94, laporte18, tepper22}), rather than self-gravity, e.g., the arches in the u-v plane \citep{minchev09,gomez12a, gomez12b}, clumps in the u-v plane \citep{quillen09}, disc asymmetries in the vertical direction resembling bending and breathing modes \citep{delavega15}. Many of these structures, however, have also been found consistent with the effect of a slow, long bar (e.g., \citealt{fragkoudi19, khoperskov2020, sanders19b, donghia20, kawata21, khoperskov22}), although it has not been easy to break the degeneracy between the tidal effect of an external perturber (e.g., Sgr) and internal perturbations from disc asymmetries, such as the bar and self-sustained spiral arms (e.g, \citealt{carrillo18, carrillo19, katz18, laport20, hunt22}).

\begin{figure*}
\includegraphics[width=14cm]{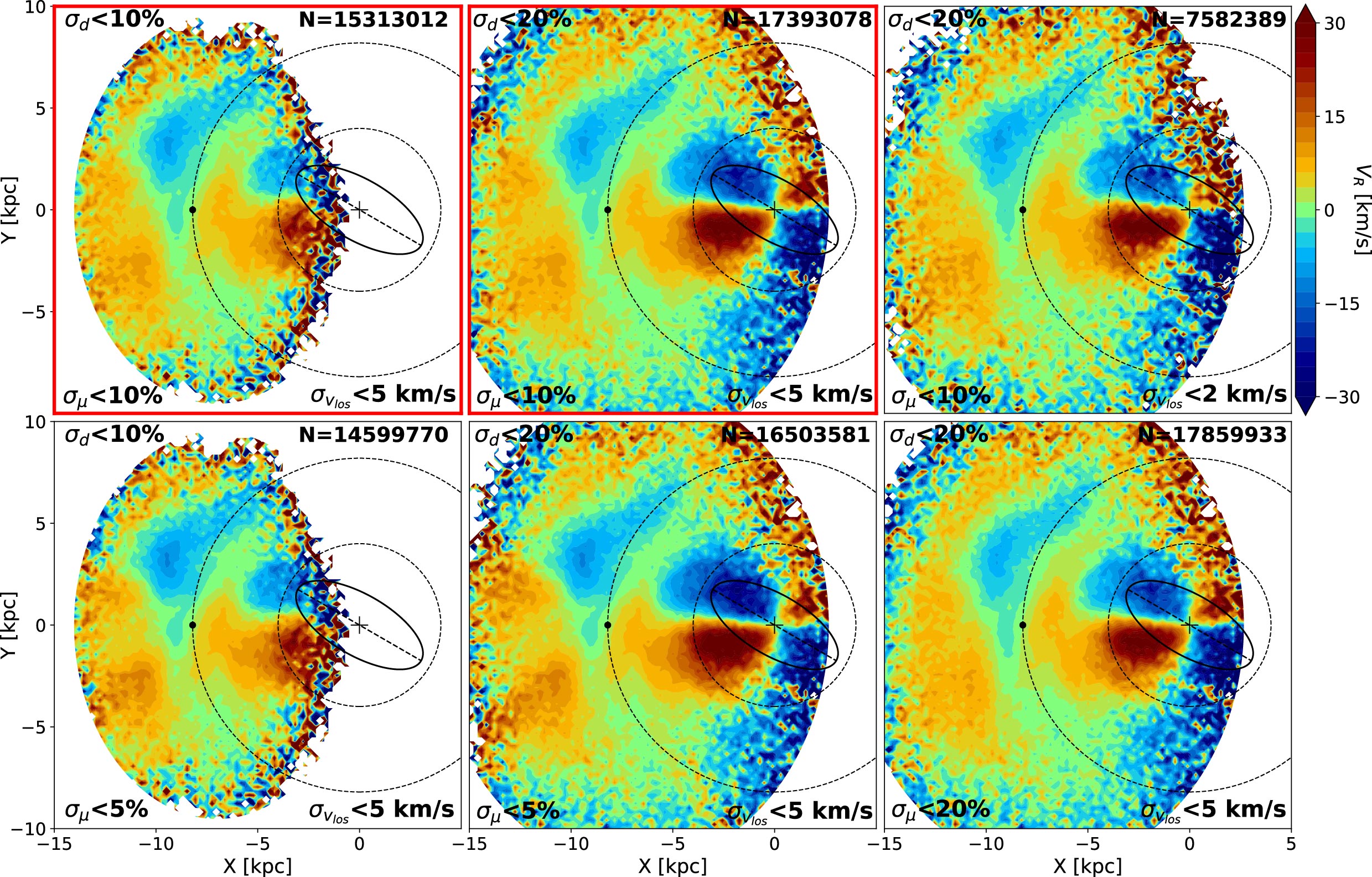}
\caption{\Gaia DR3 radial velocity field, $v_R(x,y)$ for different uncertainty cuts, as indicated. 
{Top:} $\sigma_d<10\%$ and $\sigma_{V_{los}}<5$~km/s (top left), $\sigma_d<20\%$ and $\sigma_{V_{los}}<5$~\kmps (top middle), and $\sigma_d<20\%$ and $\sigma_{V_{los}}<2$~km/s (top right). Proper motion uncertainty is $\sigma_\mu<10\%$ in all top panels.
{Bottom:} As in top, but with $\sigma_\mu<5\%$ (left and middle) and $\sigma_\mu<20\%$ (right). Significant difference in the disc area covered and the velocity map morphology are seen only when the distance measurement uncertainty changes. For comparison with our models we use the top left and middle panels. An ellipse with a semi-major axis of $3.5$~kpc depicts the bar, oriented at $30^\circ$ ahead of the Sun-GC line. The panels with $\sigma_d<20\%$ reproduce well the top-left panel of Fig.~16 by \citealt{drimmel23}, but show a larger range in $v_R$ and cover a larger disc area.
}
\label{fig:gaia}      
\end{figure*}

This drastic change in the bar length and pattern speed estimates, when measured directly from data in the inner disc, rather than modelling the local velocity field, has been rather puzzling. One way to understand this discrepancy was proposed recently by \cite{hilmi20}. These authors showed that simulated galactic bars exhibit fluctuations in length, amplitude and pattern speed, due to the periodic bar overlap with the inner spiral structure, as already noted by \cite{quillen11}. Using the ellipse-fitting ($L_{\rm prof}$) and Fourier decomposition ($L_{m=2}$) methods \citep{athanassoula02}, along with overdensity contour maps ($L_{\rm cont}$) of the central bar of MW-like simulations, \cite{hilmi20} showed that, while the bar is temporarily connected to a spiral arm it can appear up to twice its true size (see also \citealt{petersen23}), slowing down at the same time but by a smaller fraction, thus causing the ratio $\mathcal{R}=R_{CR}/R_b$ to become less than 1. This gives rise to ``ultrafast'' bars, which have been found in observations \citep{buta09, aguerri15} but theoretically deemed unphysical. Unlike the $x_1$ stellar orbits which support the bar, orbits outside the bar's CR are perpendicular to its major axis, and so $\mathcal{R}<1$ is not allowed (\citealt{contopoulos80a, contopoulos80b}). The work by \cite{hilmi20} could then explain the observed ultrafast bars with an overestimation of their length, if they happened to be connected to spiral arms. Indeed, an investigation by \cite{cuomo21} of a set of disc galaxies with ultrafast bars from the CALIFA survey, including those from \cite{aguerri15}, found that they become regular fast bars when the bar length measurement proposed by \cite{lee20} was used, based on the analysis of the maps tracing the transverse-to-radial force ratio $Q_T (r,\phi)$ of the galaxy \citep{combes81}. Another technique to overcome the biases induced by traditional bar length measurements was recently proposed by \cite{petersen23}. Dubbed 'dynamical length', this method allows to measure the extent of $x_1$ orbits, which defines the unambiguous bar length.

While the MW bar length has been estimated to be $\sim 5$~kpc or even larger (e.g., \citealt{wegg15, li16, portail17}), the more recent work by \cite{lucey23} found, via orbit integration, that trapped bar orbits extend out to only $\approx3.5$ kpc, although there is an overdensity of stars at the end of the bar, out to 4.8 kpc, which could be related to an attached spiral arm. Another recent study on exploring the bar pattern speed indirectly from the effect on the tidal stream of the Hyades \citep{thomas23} found $\Omega_b\approx55$~km/s, which is in stark contrast to the direct Tremaine-Weinberg (TW) method measurements (e.g., \citealt{bovy19, sanders19b}). Both of the above results are very much in line with the predictions by \cite{hilmi20}.

Inspired by \cite{drimmel23}, the aim of this work is to model the \Gaia DR3 radial velocity field as a function of disc position and find out what we can learn about the Galactic bar length, as well as its orientation with respect to the spiral structure. \cite{drimmel23} showed that the kinematic manifestation of the MW bar, namely the quadrupole, or butterfly-like radial velocity pattern, when the disc is viewed face-on, seems to be aligned with the Sun-GC line, implying that the bar angle, $\phi_{b}$, is close to zero. {These authors also pointed out that the apparent orientation of the quadrupole is due to distance uncertainty and, based on comparison to simulations, argued for bar angle of about $20^\circ$.}
Indeed, the consensus agrees on a tilted bar with respect to the Sun-GC line in the range 20-$30^\circ$ ahead of the Sun \citep{bland-hawthorn16}. Here, we also seek to explain this disagreement with a set of diverse hydrodynamic simulations of MW-like galaxies by accounting for uncertainties in observable measurements similar to those present in the \Gaia DR3 dataset.

\section{Gaia DR3 data selection}
\label{sec:data}

We use the third data release from the \textit{Gaia}~(ESA) space observatory to study the velocity map of the inner disc region. The unfiltered \textit{Gaia} DR3 dataset consists of over 33 million stars with 6-dimensional phase space information \citep{drimmel23}. We use multiple quality criteria to ensure the reliability of the positions and velocities of the MW stars, needed for our analysis. More specifically, we make quality cuts in the renormalized unit weight error (RUWE) $< 1.4$~\citep{recio23}, rejection of duplicated sources (determined by the \textit{Gaia} cross-matching algorithm; \citealt{torra21}), and retention of only five-parameter astrometric solutions ({\tt astrometric\_params\_solved} = 31, \citealt{recio23}). 

To calculate positions and velocities in the galactocentric rest-frame, we assumed an in-plane distance of the Sun from the Galactic centre of $8.2$~kpc, a velocity of the local standard of rest~(LSR), of $240~\kmps$ \citep{2014ApJ...783..130R}, and a peculiar velocity of the Sun with respect to the LSR, $U_\odot = 11.1$~\kmps, $V_\odot = 12.24$~\kmps, and $W_\odot = 7.25$~\kmps~\citep{2010MNRAS.403.1829S}. The heliocentric distances were taken from the \citep{2021AJ....161..147B} catalogue.

Additional quality cuts include constraints on observable velocities and the heliocentric distance, made to ensure the data are mostly free from observation biases stemming from our location in the MW disc. To find out how these cuts affect the galactocentric mean radial velocity map, $v_R(x,y)$, in Fig.~\ref{fig:gaia} we plot six panels with different combinations of cuts in the uncertainty in proper motion, $\sigma_\mu$, distance, $\sigma_d$, and line-of-sight velocity, $\sigma_{V_{los}}$. 
It can be seen in the figure that the parameter which most significantly affects both structure in velocity space and the covered disc area is the distance uncertainty, as any cuts below 20\% begin to substantially decrease the amount of data beyond the Galactic center (see left column of Fig.~\ref{fig:gaia}). To compare to our simulations, we use two distance uncertainty cut: $\sigma_d<20\%$\footnote{{When referring to uncertainty cuts, we use $\sigma_d<20\%$, etc., to mean $\sigma_d<0.2d$, etc., interchangeably within the paper.}} (as in \citealt{recio23, drimmel23}) and $\sigma_d<10\%$, which shows a different structure inside $R=4$ kpc. Looking at the various proper motion cuts (in both RA and DEC), we notice that $\sigma_{\mu}<5\%$ seems to intensify features in the radial velocity field skewed towards the solar neighbourhood, so we maintain a 10\% uncertainty bound. {Lastly, decreasing $\sigma_{V_{los}}$ from 5 to 2~\kmps makes little difference in the radial velocity field of the data \citep[see also][]{kordopatis23}, other than to slightly lessen the intensity of the butterfly pattern in the bar region, while it cuts the data sample in half. We thus maintain a 5~\kmps error limit in $V_{los}$. After all these quality cuts, we are left with a star count of about 17.4 or 15.3 million for the $\sigma_d<20\%$ and $\sigma_d<10\%$ cuts, respectively.} The variation of the radial velocity field with the above-described combinations of uncertainty cuts is shown in Fig.~\ref{fig:gaia}, as indicated in each panel.

\begin{figure*}
\includegraphics[width=15cm]{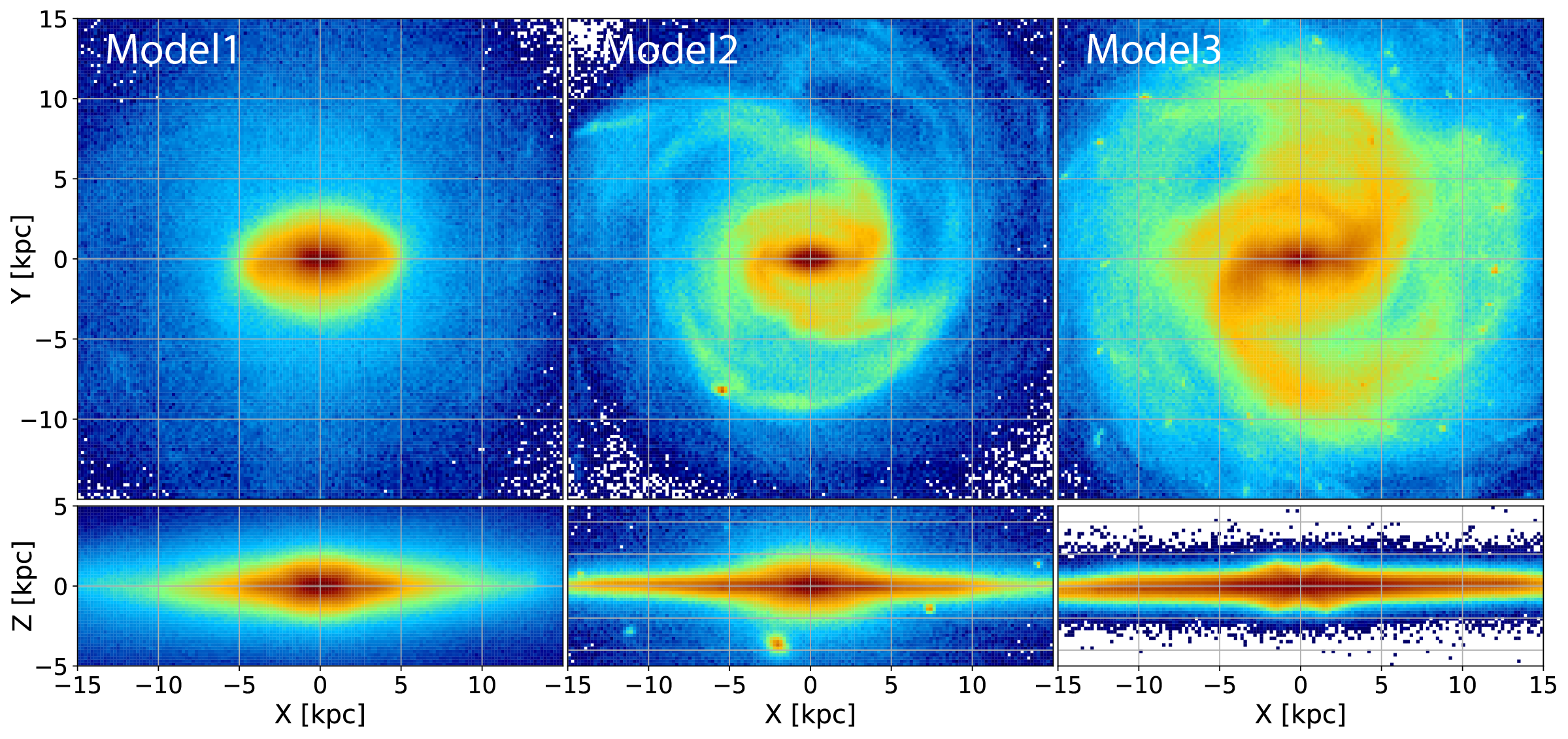}
\caption{
Face-on (top) and edge-on (bottom) stellar density plots for our Model1 (left), Model2 (middle) and Model3 (right panel). X-shaped side-on bar is seen for Model1 and Model3. which have gone through a buckling phase, in contrast to Model2, which has not.
}
\label{fig:xyz}      
\end{figure*}

\section{Simulations}
\label{sec:sims}

We use three hydrodynamical simulations of barred spiral galaxies (two in the cosmological context) with disc properties similar to those of the MW. We only consider the last $\sim1.4$ Gyr of evolution in these models, as we aim at matching the Galactic disc dynamical state at redshift zero. Since the latter is a very strong function of time, we examine closely spaced time outputs (from 4.5 to 10 Myr) in each simulation. 

Model1 and Model2, as introduced below, were used under the same names by \cite{hilmi20} to study the bar length fluctuations due to the bar-spiral interaction resulting from their different pattern speed. We keep the same names here for consistency, although the scaling we do is slightly different, as described below. Model3 is a preassembled stellar disc simulation, including gas dynamics, star formation and chemical evolution. The face-on and edge-on views of the three models for the last time outputs can be seen in Fig.~\ref{fig:xyz}. 

The angle the Galactic bar semi-major axis makes with the Sun-GC line is referred to as the bar angle, $\phi_{b}$. In our simulations, we assume a bar angle of $\phi_{b} = 30^{\circ}$, which is consistent with the upper limit of measurements found using distributions of red clump giants from surveys such as the Via Lactea, (OGLE) III, and 2MASS, e.g., $27^{\circ} \pm 2^{\circ}$ \citep{wegg13}, $29.4^{\circ}$ \citep{cao13}, and $20^{\circ} - 35^{\circ}$ \citep{lopez05}.

The velocities of our models are scaled so that the rotation curve is 240 km/s, while we scale the distances (and thus the bar lengths) such that the observed radial velocity field, $v_R(x,y)$, is reproduced as best as possible. The latter results in different bar lengths, as indicated below.

\subsection{Model1}

The first simulation (Model1) we use was introduced by \cite{buck18} as a higher-resolution of the galaxy g2.79e12 from the NIHAO project \citep{wang15}, with a notable boxy/peanut-shaped bulge similar to that of the MW \citep{buck18, buck19b}. The simulation was made with a modified version of the smoothed particle hydrodynamics~(SPH) code GASOLINE2.0 \citep{wadsley17}. The updated hydrodynamics adopt a metal diffusion algorithm between particles \citep{wadsley08}. Star formation in this model follows that described by \cite{stinson06} for dense and cool gas. Two modes of stellar feedback are used in the simulation \citep{stinson13a}; one modelled from luminous young stars before any supernovae, and the other mode from supernovae after 4 Myr of star formation.

This simulated galaxy is resolved with {$\sim5.2 \times 10^{6}$ dark matter particles ($5.141 \times 10^5 \textup{M}_\odot/$particle), {$\sim8.2 \times 10^6$ star particles} ($3.13 \times 10^4 \textup{M}_\odot/$particle, total $\rm M_{star} = 1.59 \times 10^{11} \textup{M}_\odot$), and {$\sim2.2 \times 10^6$ gas particles} ($9.38 \times 10^4 \textup{M}_\odot/$particle, total $\rm M_{gas} = 1.85 \times 10^{11} \textup{M}_\odot$) - see tables 1 \& 2 of \cite{buck19c}. }
Due to its similarity with the MW, this galaxy has been extensively studied for the birth radii of stars \citep{lu22,lu22a,lu22b,wang23}, the chemical abundance distribution \citep{buck20,sestito21,buck23} and its satellite galaxy population \citep{buck19a}.

Model1 has a flat rotation curve at $V_c = 324$ km/s, which we scaled down to match current estimates for the MW of $V_c = 240$ km/s \citep{bland-hawthorn16}. 

Unlike in \cite{hilmi20}, we do not scale the distances down, since the \Gaia DR3 radial velocity field is matched well with the original bar length, $R_b\approx5.2$ kpc, as measured from the minimum of its fluctuation in the last $\sim1.4$ Gyr (see \citealt{hilmi20} for details). The time outputs from this simulation are every 6.9 Myr.

\subsection{Model2}

The second simulation (Model2) is galaxy g106 from a suite of 33 hydrodynamic simulations by \cite{martig12} made by extracting merger and accretion histories of a particular halo from a cosmological simulation, then re-simulating with the Particle-Mesh code \citep{bournaud02, bournaud03} at high resolution with a galaxy in place of the halo (zoom-in technique introduced by \cite{martig09}). This simulation has a mass resolution of $1.5 \times 10^{4}\textup{M}_\odot$ for gas particles, $7.5 \times 10^{4}\textup{M}_\odot$ for stars, and $3 \times 10^{5}\textup{M}_\odot$ for dark matter particles; the spatial resolution is 150~pc. Within the optical radius of 25~kpc, this simulation has a total stellar mass of $\sim4.3\times10^{10}~M_\odot$ and a dark matter mass of $\sim3.4\times10^{11}~M_\odot$.
This simulation has also been studied extensively due to its similarity to the MW (e.g. \citealt{martig14a, martig14b}, \citealt{kraljic12}, \citealt{mcm13, minchev14, minchev15, minchev17, carrillo19, hilmi20}).

Similarly to \cite{hilmi20}, we scale this simulation so that the original rotation curve of $V_{\rm c}\approx210$~km/s matches the MW at 240 km/s and the bar length is fixed at $\sim3.2$ kpc at the final time, by scaling distances down by a factor of 1.46. In the 1.37 Gyr period considered, the bar length increases from $\sim2.8$ to $\sim3.2$ kpc with a time average of 3 kpc. Time outputs here are separated by 4.5~Myr. 

We re-scale Model2 distances to shrink the bar, but do not do this for Model1, in order to show that both these models, with largely differing bar lengths, can reproduce the \Gaia DR3 radial velocity field, the reason for this being the weaker spiral structure in Model1.

\begin{figure*}
\includegraphics[width=15cm]{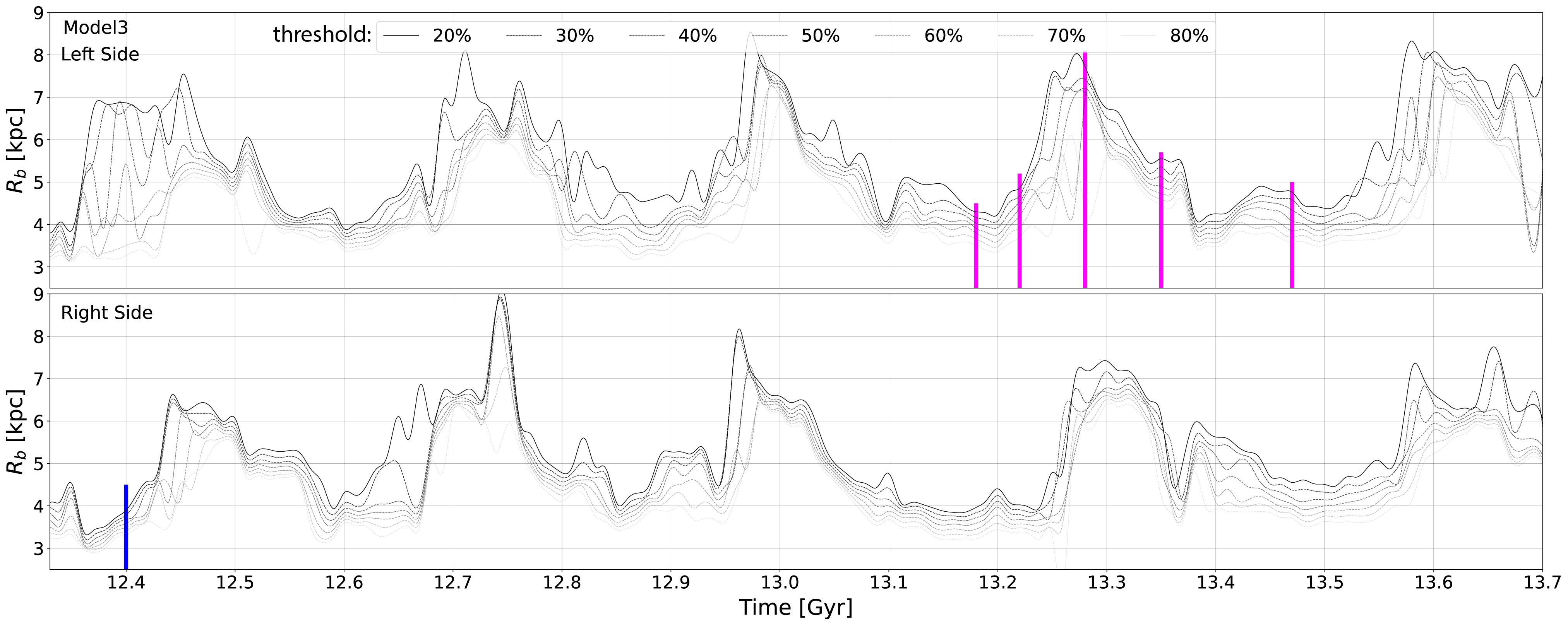}
\caption{Bar length variation for Model3, measured using the $L_{\rm cont}$ method of \protect\cite{hilmi20}. Very similar results are found with the $L_{\rm prof}$ method from, e.g.,\protect\cite{athanassoula02} (not shown). After the bar is aligned along the x-axis in a face-on view as in the top panels of Fig.~\ref{fig:xyz}, each bar half is measured separately, with the "Left Side" (extending at $x<0$) shown in the top panel and the "Right Side" (at $x>0$) shown in the bottom. The different curves represent different thresholds from 20\% to 80\% (drop in overdensity along the bar semi-major axis). The bar true length is $\sim3.6$ kpc, estimated as the minimum of the fluctuations for a 50\% threshold as done by \protect\cite{hilmi20}. Lower thresholds result in longer bar measurements, therefore 50\% is the line right in the middle for any given time. Five main peaks are seen in both the left and right bar halves (a period of $\sim270$ Myr), however, a slight offset exists, since the two bar halves do not connect to spiral arms always at the same time. The pink lines indicate the times of time outputs shown in Fig.~\ref{fig:xyt}. The blue line indicates the time output for which we find the best match to the \Gaia DR3 data (see Fig.~\ref{fig:match}). 
}\label{fig:model3_length}      
\end{figure*}

\subsection{Model3}

Our Model3 is an $N$-body/hydrodynamical simulation of a disc galaxy with a total stellar mass and a rotation curve compatible with those of the MW. The simulation starts from a pre-existing axisymmetric stellar disc, including gas and star formation coupled with chemical evolution. The simulation lasts about $3$ Gyr of which we consider the last $1.37$~Gyr as in our other two models and assume this represents well the last 1.37 Gyr of MW evolution dynamically. A well-defined buckled bar is formed before the time period we consider, as can be seen in Fig.~\ref{fig:xyz}. The detailed description of the Model3 setup is as follows:

Initially, stellar particles are redistributed following a Miyamoto--Nagai density profile ~\citep{miyamoto75} that has a characteristic scale length of $4$~kpc, vertical thicknesses of $0.2$~kpc and mass of $4.5 \times 10^{10}~\textup{M}_\odot$. Also included is a live dark matter halo~($5\times 10^6$ particles), whose density distribution follows a Plummer sphere~\citep{plummer1911}, with a total mass of $6.2\times 10^{11}~\textup{M}_\odot$ and a radius of $21$~kpc. The choice of parameters leads to a galaxy mass model with a circular velocity of $\approx220$ km/s. The gas component is represented by an exponential disc with a scale length of $5$~kpc and a total mass of $1.5 \times 10^{10}~\textup{M}_\odot$. The initial equilibrium state has been generated using the iterative method from the AGAMA software~\citep{vasiliev19}. A gaseous cell undergoes star formation if: i) the gas mass is $> 2\times10^5 \textup{M}_\odot$, (ii) the temperature of the gas is lower than $100$~K and (iii) if the cell is part of a converging flow. The efficiency of star formation is set to $0.05$, i.e., $5\%$ of the gas eligible to form a new star particle per dynamical time.  We consider the ISM as a mixture of several species (H, He, Si, Mg, O, Fe, and other metals), which is sufficient for modelling the Galactic chemical evolution and the newborn stellar particles inherit both kinematics and elemental abundances of their parent gas cells. No chemical information is used in this work.

Following the chemical evolution models by~\cite{snaith15} and \cite{2022A&A...659A..64S}, at each time step, for newly formed stars we calculate the amount of gas returned, the mass of the various species of metals, the number of SNII or SNIa for a given initial mass and metallicity, the cumulative yield of various chemical elements, the total metallicity, and the total gas released. Feedback associated with the evolution of massive stars is implemented as an injection of thermal energy in a nearby gas cell proportional to the number of SNII, SNI and AGB stars~\citep[see][for more details]{khoperskov21}. The hydrodynamical part also includes gas-metallicity-dependent radiative cooling~\citep[see details in][]{khoperskov21}. The simulations were evolved with the $N$-body+Total Variation Diminishing hydrodynamical code~\citep{khoperskov14}. For the $N$-body system integration and gas self-gravity, we used our parallel version of the TREE-GRAPE code~\citep[][]{fukushige05} with multithread usage under the SSE and AVX instructions. In recent years we already used and extensively tested our hardware-accelerator-based gravity calculation routine in several galaxy dynamics studies where we obtained accurate results with a good performance~\citep{khoperskov18a,khoperskov18b,saburova18,khoperskov19}. For the time integration, we used a leapfrog integrator with a fixed step size of $0.1$~Myr. In the simulation, we adopted the standard opening angle $\theta = 0.7$. The dynamics of the ISM is simulated on a Cartesian grid with static mesh refinement and a minimum cell size of $\approx10$~pc in the Galactic plane.

Similarly to Model1 and Model2, we scale this simulation's rotation curve from 220 km/s to 240 km/s and do not scale the distances. The bar length during the period of time we consider is $\sim3.6$ kpc, as estimated from the minima of the 50\% threshold in Fig.~\ref{fig:model3_length}. As in Model1 and Model2, we chose this bar length as it happens to match well the \Gaia DR3 radial velocity field. Time outputs here are separated by 10~Myr. 

\subsection{Spiral structure}
\label{sec:sp}

The spiral arms of Model1 are more tightly wound and multi-armed (see Fig.~\ref{fig:xyz} and Fig.~1 by \citealt{buck18}), compared to Model2 or Model3, where they are more open and dominated by two or four arms (see also Fig.~1 by \citealt{mcm13}), signifying that they are stronger. We measured the spiral structure overdensity from the ratio of the amplitude of the m=1,2,3 and 4 components to the m=0 Fourier component of the stellar density, as a function of galactic disc radius and that for three radii in Fig.~\ref{fig:fourier}. We found that the spiral overdensity is typically 10\% for Model1 and about 20-25\% Model2 and Model3. {One important difference is that Model3 displays the strongest odd modes: m=1 and m=3, which correspond to a one- and three-armed spirals (see Fig.~\ref{fig:fourier}).}

For the MW we expect spiral-arm overdensity of $\sim15\%$ from modelling the radial velocity field of RAVE data \citep{siebert12}, $\sim25\%$ from considering the migration rate of open clusters near the Sun \citep{quillen18a}, $\sim20\%$ from matching the radial velocity field of stars from a compilation of data \citep{eilers20}. These estimates are larger than the spiral overdensity of Model1, but consistent with our Model2 and Model3.

\subsection{Matching the selection and uncertainties of \Gaia DR3}
\label{sec:err}

In order to compare properly to the observations, we need to match the geometry of the \Gaia DR3 sample and to introduce synthetic uncertainties into our models, consistently with the data. 

We first transform our simulation data from the native galactocentric cylindrical coordinate system to a Galactic spherical coordinate frame centred on the Sun. This is done with the {\tt astropy.coordinates} module using the {\tt SkyCoord.transform\_to} object method (see \citealt{astropy2022}). In this transformation we specify the position of the Sun's barycentre at $(x,y,z)=(-8.2,0,0)$~kpc in the galactocentric frame, as this is shifted to the origin of the new Galactic frame. To match the reference frame from which \Gaia measures kinematic observables, we perform another simple coordinate transformation of the simulated data, taking it from Galactic spherical to ICRS coordinates.

To match the geometry of our \Gaia sample, we picked the Sun position in the disc so that the bar is oriented at $30^\circ$ ahead of the Sun-galactocentric line \citep{bland-hawthorn16}, {but we test the case of $20^\circ$ as well.} Moreover, we made a cut in Galactic latitude, $|b|<1.0^\circ$ about the midplane, in order to exclude the dust-obscured regions in the data and match the \Gaia footprint. This achieves a similar effect as our quality cuts in the \Gaia DR3 data, which preferentially reject stars within the Galactic midplane.

{The \Gaia DR3 uncertainties for our dataset revealed roughly Gaussian distributions in proper motion and line-of-sight velocity uncertainties and a complex, skewed Gaussian distribution in distance uncertainty which is coupled to distance, as shown in the left panel of \ref{fig:errMod}. This skewing towards larger distances was already shown by \cite{drimmel23}. As for the data, we introduced synthetic uncertainties in the line-of-sight velocity {$\sigma_{V_{los}}=5$~km/s}, in heliocentric distance $\sigma_d=0.2d$ or $\sigma_d=0.1d$, and in both proper motions $\sigma_\mu=0.1\mu$, using {Gaussian} distributions. To model the distribution of relative uncertainties in distance ($\sigma_d\slash d$) in two different ways. First we use a Gaussian distribution of width 20\% or 10\% {centered on zero}, which is used for most of the figures. We also fit the data using {skewed probability distributions fitting the data in different distance bins}. These functions are then interpolated to create a continuous probability density function dependent on distance. The result is shown in the right panel of Fig.~\ref{fig:errMod} and provides a very good match to the data on the left.} After convolving the uncertainties, we converted from ICRS back to Galactocentric cylindrical coordinates. Now our models include the biases in the \Gaia DR3 observables and can allow for proper comparison to the data.

\begin{figure}
\includegraphics[width=8cm]{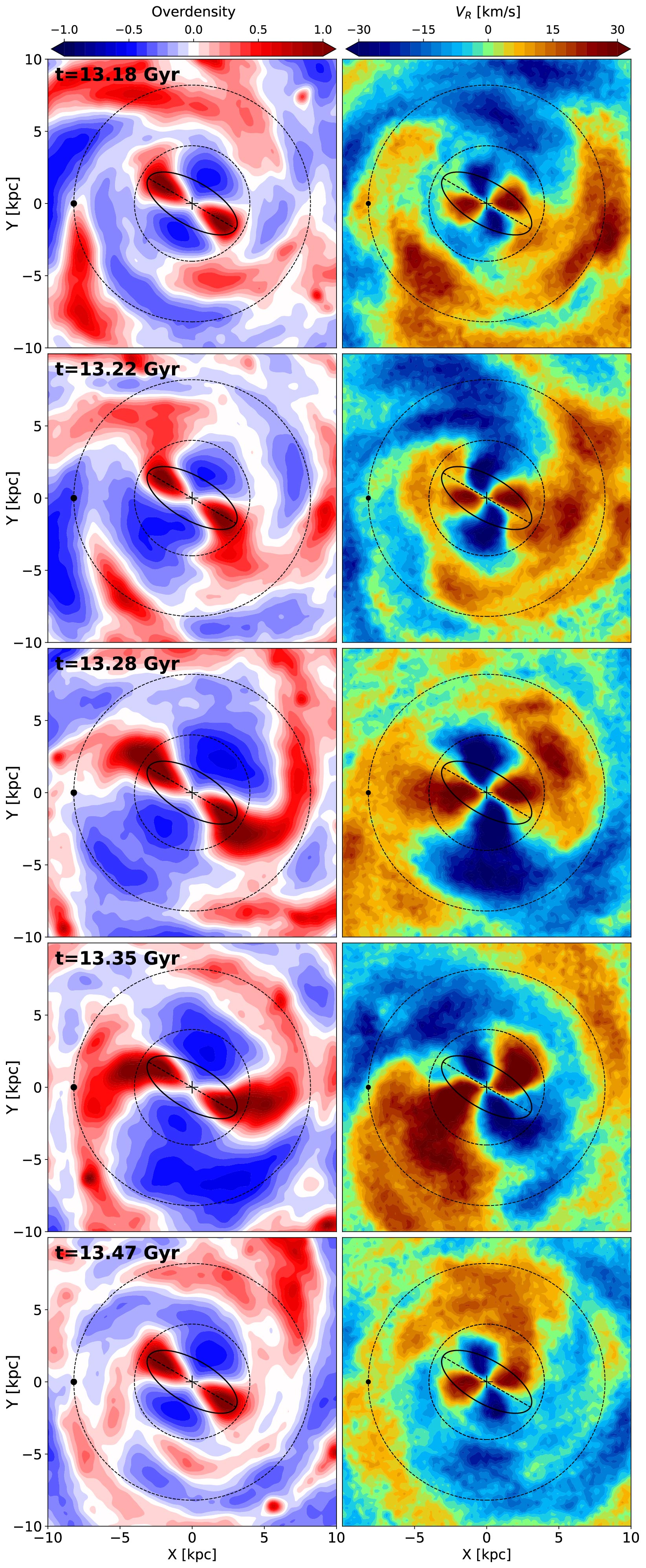}
\caption{
Illustrating the changes to the Model3 disc morphology (left column) and its radial velocity field (right column) when the bar is disconnected from the inner spiral structure (top row), in the process of connecting (second row), fully connected and at a maximum measurable length (third row), in the process of disconnecting from spiral (fourth row), and disconnected once again (bottom row). Drastic differences are seen among different panels over these very short time intervals (40-120 Myr).
 In all the plots the bar is oriented at $30^\circ$ with respect to the Sun-galactocentric line. The Sun's location is indicated by the black dot at $x=8.2$~kpc and $y=0$. The two dotted circles show the solar radius and 4~kpc.
}
\label{fig:xyt}      
\end{figure}

\begin{figure*}
\includegraphics[width=14cm]{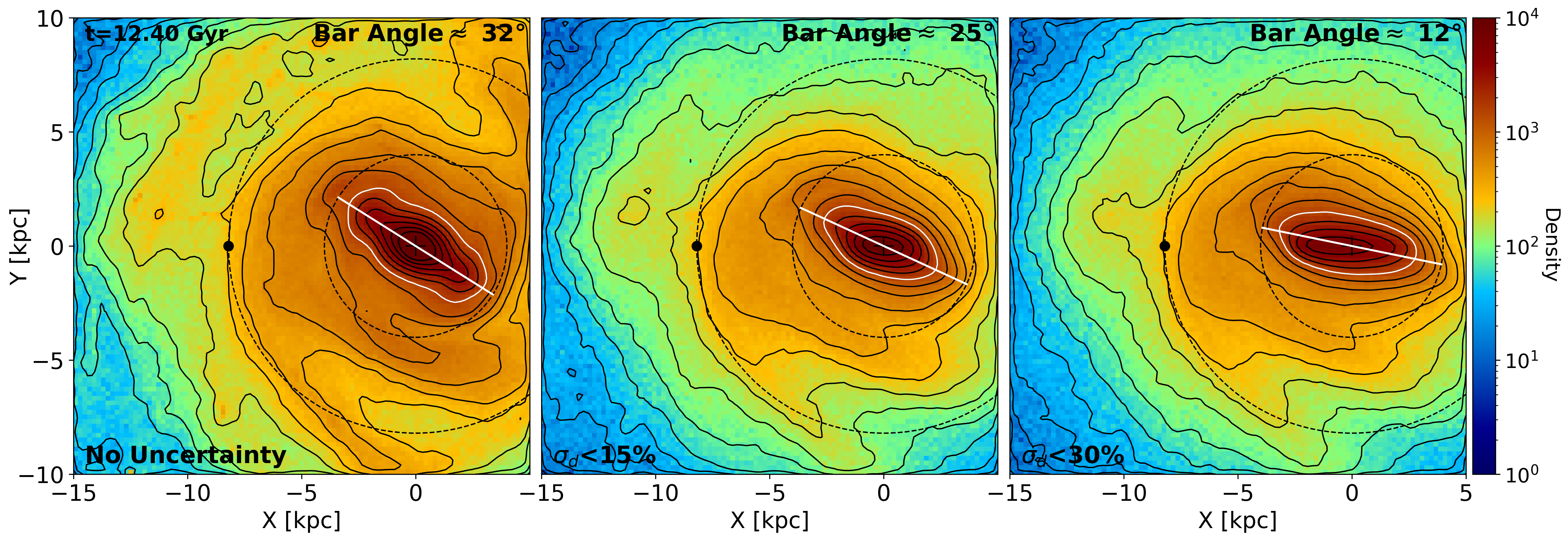}
\caption{
The effect of \Gaia DR3-like uncertainties on the bar shape and orientation. The left panel shows the face-on stellar density of Model3 at $t=12.4$~Gyr (bar length near a peak, see Fig.~\ref{fig:model3_length}) with the bar oriented at $30^\circ$ ahead of the Sun-galactocentric line. The middle and right panels show the effect of 15\% and 30\% distance uncertainty, respectively. The measurement errors, most importantly the distance uncertainty, cause the bar angle to shift significantly towards the Sun-galactocentric line, as indicated at the bottom of each panel. Moreover, the central contours are affected more, resulting in a distorted bar shape. The white contour marks the same density level, which changes from the 6th, to the 5th, to the 4th most dense contour for a $\sigma_d=0, 20\%, 30\%$, respectively. This is caused by the central density spread out due to the distance uncertainty increase. We measured {$25^\circ$ for $\sigma_d<15\%$ and $12^\circ$ for $\sigma_d<30\%$ taking all density inside the bar length (3.6 kpc) into account}. However, it is clear from the innermost couple of contours that the effect is even larger there due to the more circular initial distribution. 
}
\label{fig:angle}      
\end{figure*}

\section{Results}

\subsection{Gaia DR3 radial velocity map}
\label{sec:rad}

It has been previously shown (e.g., \citealt{carrillo19, fragkoudi19, bovy19}) that a central bar produces a quadruple pattern in the disc radial velocity field when viewed face-on. This signature was first identified for the MW by \cite{bovy19} and then more clearly by \cite{queiroz21}, combining APOGEE \citep{majewski17} spectroscopy with earlier \Gaia DRs astrometry, using a few tens of thousands of stars. The advent of \Gaia DR3 confirmed the existence of such a kinematic pattern using millions of stars, and extended to spiral arms all the way outside the solar circle. 

Fig. \ref{fig:gaia} presents the \Gaia DR3 radial velocity field, $v_R(x,y)$ for different uncertainty cuts, as indicated at the top of each panel. The panels with $\sigma_d<20\%$ (all except for the top left one) reproduce well the top-left panel of Fig.~16 by \cite{drimmel23}, but show a larger range in $v_R$ (colour bar) and cover a larger disc area. This allows us to see better the positive and negative velocity lobes on the other side of the GC, the emergence of an additional arm-like feature in negative velocity (blue) at the upper-left quadrant of the plots, and an area of negative radial velocity in the lower-right corner. An ellipse with a semi-major axis of $3.5$~kpc depicts the MW bar in the figure, assumed to be $30^\circ$ ahead of the Sun-GC line (e.g., \citealt{bland-hawthorn16}). 

The expected orientation of the bar density (ellipse in Fig. \ref{fig:gaia}) should lie along the line delineating negative from positive $v_R$ lobes, as shown in Fig.~\ref{fig:xyt}. However, we find that the bar semi-major axis is aligned with the blue lobe instead. We will show later that this is the effect of the distance uncertainty. Note that once we cut at $\sigma_d<10\%$, the inner $\sim2$ kpc velocity structure aligns better with the bar major axis.

\subsection{Rapid variations in Galactic disc morphology and radial velocity field}
\label{sec:time}

When more than one perturber is present in a galactic disc, such as a central bar and spiral arms moving at different pattern speeds, one expects a strong variation of both the density and the velocity field on short timescales~(e.g., \citealt{carrillo19,tetsuro22}). Indeed, using Model1 and Model2, \cite{hilmi20} showed that the bar length, amplitude, and pattern speed can all fluctuate on a dynamical timescale consistent with the beat frequency between the bar and inner spiral modes. 

In Fig.~\ref{fig:model3_length} we show the bar length evolution with time for our Model3. The two half lengths are measured separately with the one near the Sun shown in the top panel (Left Side) and the one farther from the Sun in the bottom (Right Side). About 5 fluctuations are seen from the number of peaks and troughs. The length measurement method used is $L_{\rm cont}$, tracing the drop in overdensity along the bar major axis, as described by \cite{hilmi20}, who reasoned that the minimum in the 50\% threshold in the density drop was closest to the true bar length, which happens when the spiral is fully disconnected from it. We see that within this definition, the bar fluctuates between $\sim 3.6$ to $\sim 7$~kpc in length, with a period of about 250-300 Myr (the beat frequency between the bar and the dominant inner spiral mode). 

To understand how the bar length fluctuations seen in Fig.~\ref{fig:model3_length} affect the inner disc morphology and its radial velocity field, in Fig.~\ref{fig:xyt} we plot the disc face-on view for five snapshots from our Model3. Those are separated by 40-120~Myr and are picked according to the relative orientation between the bar and spiral. The left column shows the stellar overdensity, computed as $\delta\Sigma(r,\phi) = (\Sigma(r,\phi) - \Sigma_0(r))/\Sigma_0(r)$, where $\Sigma(r,\phi)$ is the density as seen in the top panel of Fig.~\ref{fig:xyz} and $\Sigma_0$ is the azimuthally averaged density for radial bins of 0.3 kpc. In the right column, we plot the galactocentric radial velocity field, $v_R(x,y)$. In all panels, the bar is oriented at $30^\circ$ with respect to the Sun-GC line for a Sun position indicated by the black dot at $x=8.2$~kpc and $y=0$. The two dashed circles show the solar radius and 4~kpc to guide the eye. We can see that in different rows, spirals have different orientations with respect to the fixed bar, due to their lower pattern speed. In the reference frame of the bar, spirals move counterclockwise, although Galactic rotation is clockwise. Note that the \Gaia DR3 uncertainties and sample selection, as described in Sec.~\ref{sec:err}, are not applied to Fig.~\ref{fig:xyt}.

The time outputs shown in Fig.~\ref{fig:xyt} span about 290 Myr, starting and ending with a complete separation between the near bar half and the spiral arms. This corresponds to one period of the bar length fluctuations, seen in Fig.~\ref{fig:model3_length} (pink vertical lines). From top to bottom, the bar half nearer the Sun is well separated from the spiral arm and thus at a minimum in Fig.~\ref{fig:model3_length}, in the process of connecting (second row), fully connected (third row) and thus a maximum in Fig.~\ref{fig:model3_length}, in the process of disconnecting (fourth row), and again fully disconnected (bottom row). 

It is easy to infer from both the overdensity and $v_R$ plots the times for which the bar is separated from the spiral. In the top and bottom panels, the bar fits well within the 4-kpc dashed circle, while in the third row (fully connected) it is extending well beyond it and the positive $v_R$ lobe covers roughly four times larger area.

The above-described variations outside the bar region in both density and velocity mean that comparison between the \Gaia DR3 data and models should be done carefully, studying the detailed time evolution of the disc. To accomplish this, for all our models we use time outputs between 4.5 and 10 Myr, depending on the simulations (see Sec.~\ref{sec:sims}).

\begin{figure*}
\includegraphics[width=18cm]{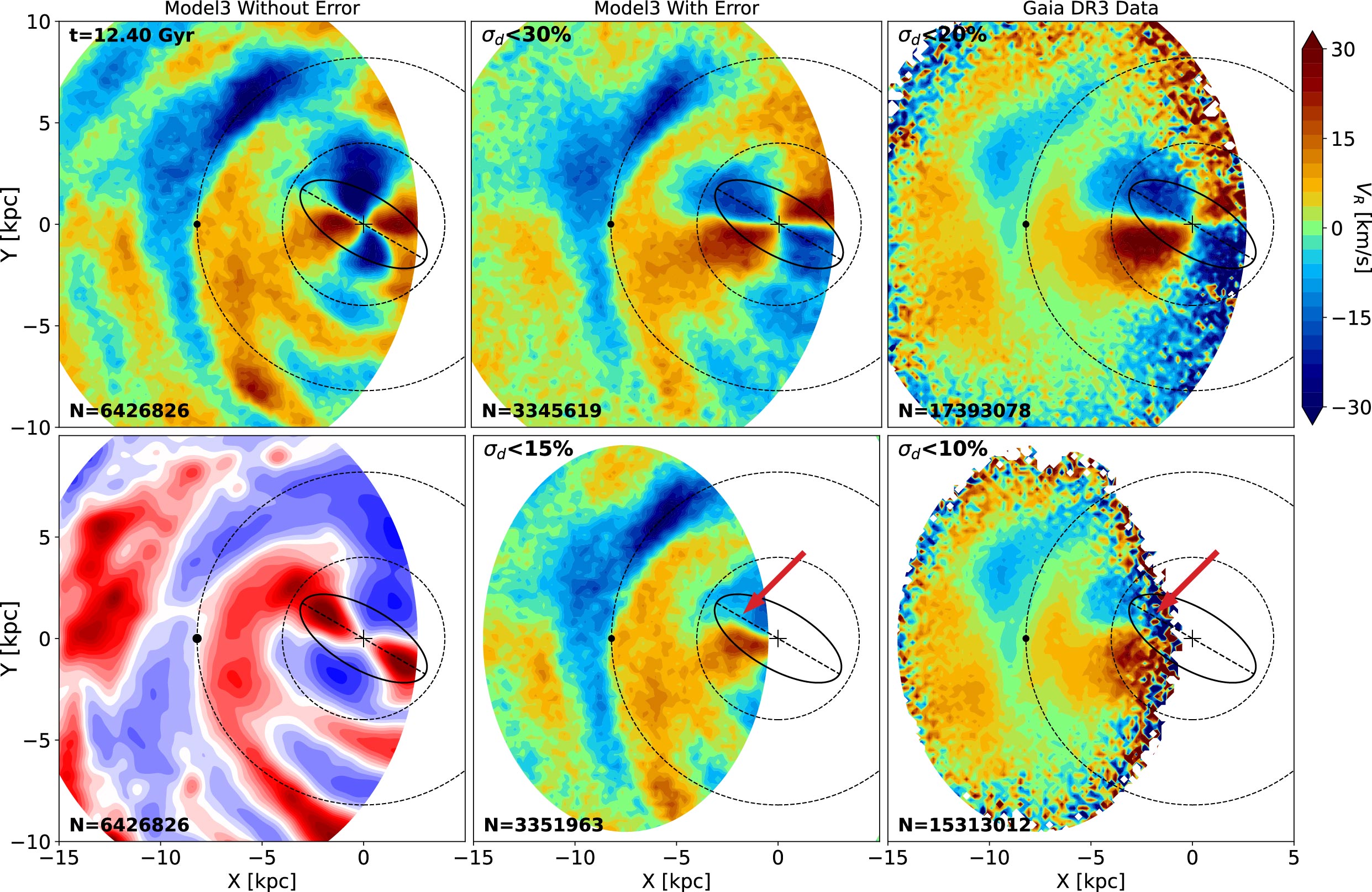}
\caption{
A snapshot from our Model3, exploring the effect of \Gaia DR3 uncertainty and providing a match to the \Gaia DR3 radial velocity field. 
{Left column:} Model3 radial velocity field (top) and stellar overdensity (bottom). {Middle column:} As top left, but including 30\% (top) and 15\% (bottom) synthetic uncertainty in heliocentric distance, $\sigma_d$.
{Right column:} Radial velocity field of \Gaia DR3 data with $\sigma_d<20\%$ (top) and $\sigma_d<10\%$ (bottom). The middle column provides an excellent match to the data. Examining Fig.~\ref{fig:model3_length}, the time 12.4 Gyr (blue vertical line) corresponds to an increase in the bar's length, thus, the bar can be thought of as being in the process of connecting to a spiral arm. Looking at the morphology of the overdensity plot (bottom-left panel), however, it appears that the bar is in the process of disconnecting from an arm (see Sec.~\ref{sec:modes} for discussion on this conundrum in terms of material arms versus spiral density wave modes). {The red arrows in the bottom middle and right panels indicate a feature, where the semi-major axis of the oval cleanly separates positive and negative velocities before it
flattens sharply at $\sim2$~kpc from the Galactic centre in both model and data.}
}
\label{fig:match}      
\end{figure*}

\subsection{Spiral arm in stellar mass near bar end due to overlap of multiple modes}
\label{sec:modes}
{
To first order, when the bar is in the process of connecting to, or disconnecting from, a spiral arm (second and fourth rows in Fig.~\ref{fig:xyt}, respectively), a leading or trailing arm, respectively, appears attached to the bar. Overall, the velocity field seen in the right panels shows similar morphology in the positive $v_R$ nearby lobe. }

{
It should be kept in mind, however, 
that multiple spiral modes with different multiplicity (typically m=1,2,3, and 4) and different pattern speeds, are always present just outside the bar, as seen in both numerical simulations (e.g., \citealt{sellwood88, quillen11, minchev12a, hilmi20}) and observations (e.g., \citealt{elmegreen92, rix93, henry03, meidt09}). This has been shown to be the case also for Model1 and Model2 by \cite{hilmi20}, by constructing power spectrograms. Therefore, the spiral overdensity seen in the mass in the left column of Fig.~\ref{fig:xyt} is due to the overlap of all these modes and not caused by a single spiral pattern (although dominated by the strongest mode). It is thus possible that while the bar appears to be connecting to, or disconnecting from this apparent spiral (which, in fact, is an overdensity associated with the overlap of the multiple modes), signatures of both disconnecting and connecting spiral arms are present in the $v_R$ field in the same snapshot. Indeed, in the second row of Fig.~\ref{fig:xyt} we can see a trailing arm in positive $v_R$ (right column) just outside the 4-kpc dashed circle, in addition to the leading arm seen in the density we reported above, although we see no spiral overdensity in the left panel.
}

{We estimated the m=1,2,3, and 4 Fourier components from the face-on density of each model as functions of time with the results shown in Fig.~\ref{fig:fourier}. It can be seen there that as stated before, Model1 has significantly weaker spiral structure than the other models. While Model2 and Model3 have similar 2- and 4-armed modes, Model3 has stronger m=1 and m=3 components overall. We later argue that these odd modes are not expected to be very strong for the MW. }

\begin{figure*}
\includegraphics[width=16cm]{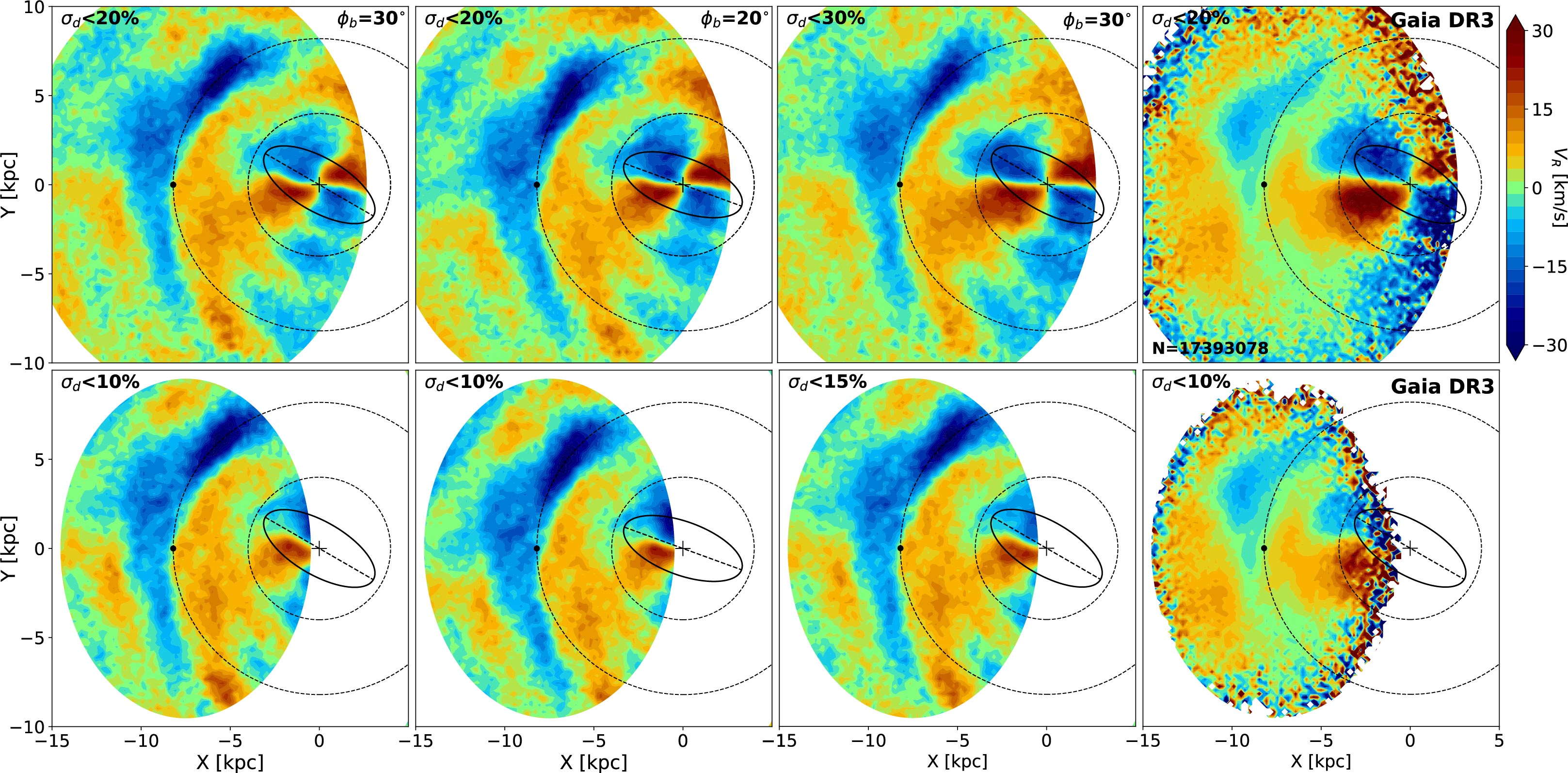}
\caption{
Exploring the interplay between distance uncertainty and bar angle. The left three columns show the Model3 radial velocity field, $v_R(x,y)$, for the matching snapshot at 12.4 Gyr (see Fig.~\ref{fig:match}), but with different combinations of distance uncertainty and bar angle, as indicated. When the implemented distance uncertainty is 20\% and the bar angle is 30$^\circ$ (leftmost column), as in the data (rightmost column), an upward kink inside the oval representing the bar is seen for the positive $v_R$ lobe, which is not present in the data. To achieve the flatness of the transition between negative and positive $v_R$, we propose two solutions: decreasing the bar angle from 30$^\circ$ to 20$^\circ$, while setting the distance error at 20\% (second column), or using an uncertainty of 30\% and a bar angle of 30$^\circ$ (third column, as in Fig.~\ref{fig:match}). The latter solution appears to give a better match to the data (rightmost column), suggesting that \Gaia DR3 distance uncertainties are underestimated.
}
\label{fig:gaia_comp}      
\end{figure*}

\subsection{Effect of \Gaia DR3 uncertainties on the bar shape and orientation}

We now study the effect of \Gaia DR3-like uncertainties on the disc morphology of our Model3, but the results are very similar for our other two Models. Fig. \ref{fig:angle} shows the face-on stellar density when no uncertainties are considered (left), for an adopted distance uncertainty of $\sigma_d<15\%$ (middle), and $\sigma_d<30\%$ (right). 

It is immediately obvious that the bar angle is strongly decreased from its true $30^\circ$ when synthetic uncertainties are included in the simulation, as already expected from \Gaia mock catalogues \citep{romero15} {and the work by \cite{drimmel23}}. We measured $\sim25^\circ$ for $\sigma_d<15\%$ and $12^\circ$ for $\sigma_d<30\%$ {taking all density inside the bar length (3.6 kpc) into account}. However, it is clear from the innermost couple of contours that the effect is even larger there (close to zero in the right panel), due to the more circular initial distribution. 
{The white contour in each panel marks the same density level, which shifts from the 6th, to the 5th, to the 4th most dense contour with an increase in uncertainty: $\sigma_d=0, 20\%, 30\%$, respectively. This indicates that the central density spreads out as the distance uncertainty increase. This stretch in the initially circular central density contour can be also seen in the stellar velocity dispersion, as shown by \cite{hey23}.}

In addition to the decrease in bar angle, the uncertainties cause the bar to appear less centrally concentrated and offset from the galactic centre toward the direction of the Sun, as seen in the figure (especially for the 30\% error). It is notable that the contours that encapsulate the bar are affected differently by $\sigma_d$. The highest density contour is almost aligned with the Sun-GC line for both distance uncertainty cuts, which can be linked to its originally nearly circular shape. 

As it can be already expected, we show in the next section that this apparent decrease in the bar orientation strongly affects the observed orientation of the central radial velocity field, $v_R(x,y)$, as well.

\subsection{Matching the \Gaia DR3 radial velocity field}

Fig.~\ref{fig:match} presents a snapshot at $t=12.4$ Gyr from our Model3, exploring the effect of \Gaia DR3 uncertainties and providing a match to the disc radial velocity field. The left column shows the model $v_R(x,y)$ (top) and the stellar overdensity (bottom), with no uncertainties included. Since this is a snapshot when the near half of the bar is connected to a spiral, the bar appears much longer than its true length, which at this particular time is $R_b\approx3.2$ (see minimum at $t\approx12.37$~Gyr in Fig.~\ref{fig:model3_length}). Using the ellipse fitting method, $L_{\rm prof}$, we measured $R_b\approx5.5$~kpc apparent bar length, however, due to the gap present in the stellar overdensty along the bar semi-major axis (bottom left panel of Fig.~\ref{fig:match}), the $L_{\rm cont}$ method introduced by \cite{hilmi20}, measured a lower value, more consistent with the real length. It should be kept in mind that observationally  $L_{\rm cont}$ cannot be applied to the MW,\footnote{{The $L_{\rm cont}$ method uses the disc overdensty, which requires knowledge of the correct Galactic density as a function of position and for all Galactic azimuths in the inner 5-6 kpc (see \citealt{hilmi20}).}} thus a bar at this configuration will be likely miss-measured by a factor of $\sim1.7$. This also results in a larger positive radial velocity lobe, compared to when the bar is disconnected from the spiral structure, as it was already illustrated in Fig.~\ref{fig:xyt}.

The middle column of Fig.~\ref{fig:match} shows the radial velocity field as in top left, but including 30\% (top) and 15\% (bottom) synthetic error, in addition to the uncertainties in heliocentric radial velocity, $\sigma_{V_{los}}$, and proper motion, $\sigma_\mu$ (see Sec.~\ref{sec:data}). As expected from the decrease in bar angle caused by the uncertainty that we saw in Fig.~\ref{fig:angle}, the butterfly pattern in the centre is rotated counterclockwise so that the bar semi-major axis (dashed line) passes through the negative velocity lobe (blue), instead of the interface between positive and negative (as in top left). We note that our results of the bar angle decreasing with {distance} error are in agreement with conclusions by \cite{drimmel23} and \cite{hey23}. The more important effect of the relative bar-spiral orientation, however, has not been discussed before.

\begin{figure*}
\includegraphics[width=18cm]{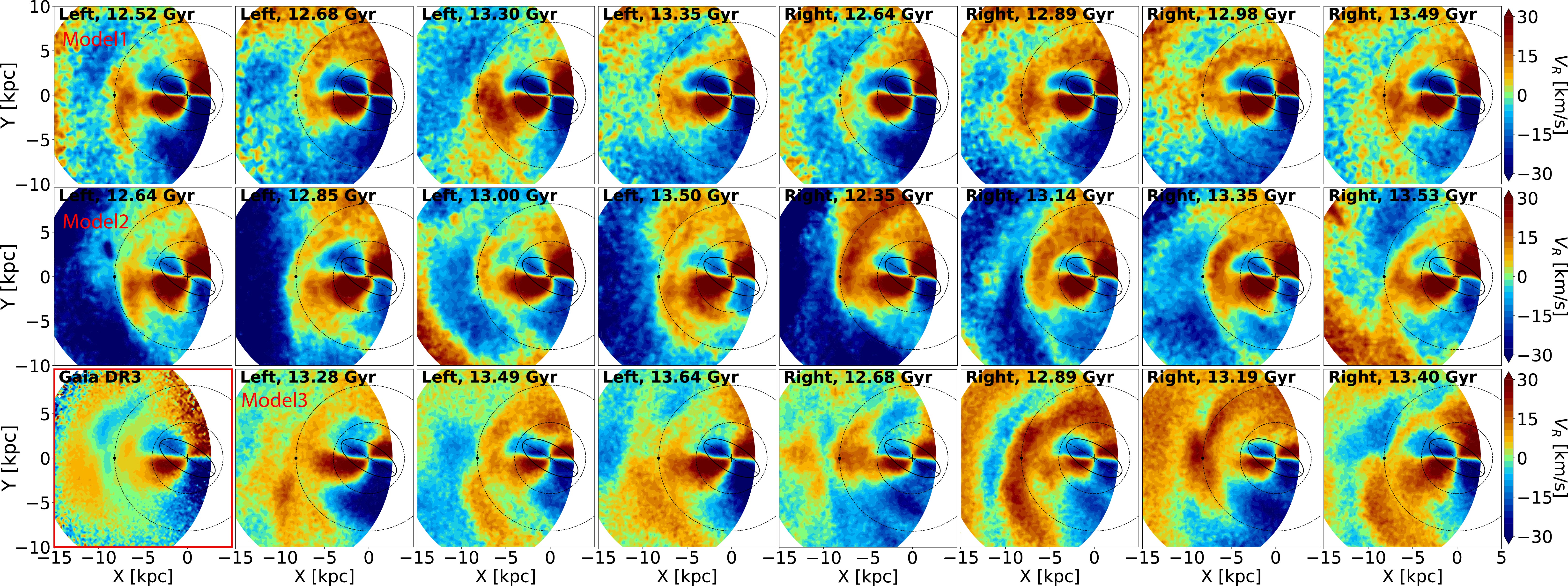}
\caption{
A selection of matches to the \Gaia DR3 radial velocity field (shown in bottom left panel) from our three different simulations, by visual inspection, considering different time outputs and focusing mostly on structure inside the solar circle. The top row shows Model1 with a true bar length of $\sim5.2$ kpc, but weaker spiral arms. The second row shows Model2, which has a true size of $\sim3$ kpc (varying from $\sim2.8$ to $\sim3.2$ kpc over the examined time period). The bottom row shows Model3, with a true bar length of $\sim3.6$ kpc, which provided the best match the the \Gaia DR3 data in Fig.~\ref{fig:match} (we do not repeat that time output here). For all snapshots we used the same synthetic uncertainties as in Fig.~\ref{fig:match}.
It is remarkable that this large range of bar lengths (from $\sim3$ to $\sim5.2$ kpc) among our three models can match relatively well the structure inside the solar circle. We attribute this mostly to the relative overdensity between the bar and spiral arms - the stronger the spiral arms, the smaller the bar that can reproduce the observations.
}
\label{fig:matches_all}      
\end{figure*}

Finally, in the right column of Fig.~\ref{fig:match} we show the radial velocity field of \Gaia DR3 data with $\sigma_d<0.2d$ (top) and $\sigma_d<0.1d$ (bottom). The middle column provides an excellent match to the entire \Gaia DR3 dataset, especially inside the solar circle in the following:
\begin{itemize}
\item the sizes and shapes of the negative and positive radial velocity lobes associated with the bar's near half;
\item the upward extending positive $v_R$ arm attached to the positive $v_R$ lobe;
\item the innermost 2-3 kpc for the uncertainty in data of 10\% (15\% in model, see Sec.~\ref{sec:inter}), where the semi-major axis of the oval (3.5 kpc long) cleanly separates positive and negative velocities before it flattens sharply at $\sim2$~kpc from the Galactic centre {(see red arrows)}; 
\item the sharp decrease in positive $v_R$ area below the positive $v_R$ lobe;
\item the wide positive $v_R$ arm along the left side of the plot, with a bifurcation in the upper left quadrant.
\end{itemize}

{The reason in Fig.~\ref{fig:match} we compared data with 10\% and 20\% distance uncertainty cut to 15\% and 30\% in the model is because increasing the distance uncertainty was the only way to achieve the flatness of the transition between negative and positive $v_R$ lobes in Fig.~\ref{fig:match} (see Sec.\ref{sec:inter} and Fig. \ref{fig:gaia_comp} for the effect of 10\% and 20\% uncertainties). This suggests that the data distance uncertainties are underestimated, but see the next section for more discussion on this.}

\subsection{Interplay between distance uncertainty and bar angle}
\label{sec:inter}

In Fig.~\ref{fig:match} we showed a time output from our Model3 simulation, which provided an excellent match to the \Gaia DR3 data. However, the implemented distance uncertainty in the simulation was 15\% and 30\%, rather than the 10\% and 20\% cuts in the data. To justify this, in the top left panel of Fig.~\ref{fig:gaia_comp} we show the same Model3 snapshot but with 10\% and 20\%. It can be seen that the flatness of the transition between negative and positive $v_R$ cannot be achieved with the 20\% error, although the difference between 10 and 15\% is not so dramatic. Therefore, an uncertainty of 30\% needs to be used, as we did in Fig.~\ref{fig:match}, to match the data cut of 20\%. 

We also considered the possibility that the bar angle is smaller than $30^\circ$, which would then require smaller uncertainty in the simulation to align the bar with the Sun-GC line. In Fig.~\ref{fig:gaia_comp} we explore how well this particular Model3 snapshot (at 12.4 Gyr) matches the data when the implemented error is 20\% as in the data, but changing the bar angle from 30 to $20^\circ$, which can be seen as a lower limit \citep{bland-hawthorn16, drimmel23}. 

Indeed, we can see that, even though $\sigma_d=20\%$ combined with $30^\circ$ (top left panel) does not provide a good match to the data (top rightmost panel), when the angle goes down to $20^\circ$ the comparison with \Gaia DR3 is much better, though arguably not as good as in the top-third panel (same as the top-middle panel of Fig.~\ref{fig:match}. 

{Finally, using the Gaussian distance uncertainty modeled as in the data (Fig.~\ref{fig:errMod}), we show in Fig.~\ref{fig:match2} that we also require a bar angle of $20^\circ$ to match well the \Gaia data. More work is needed to explore the interplay between bar angle and distance uncertainties.}

\begin{figure*}
\includegraphics[width=17cm]{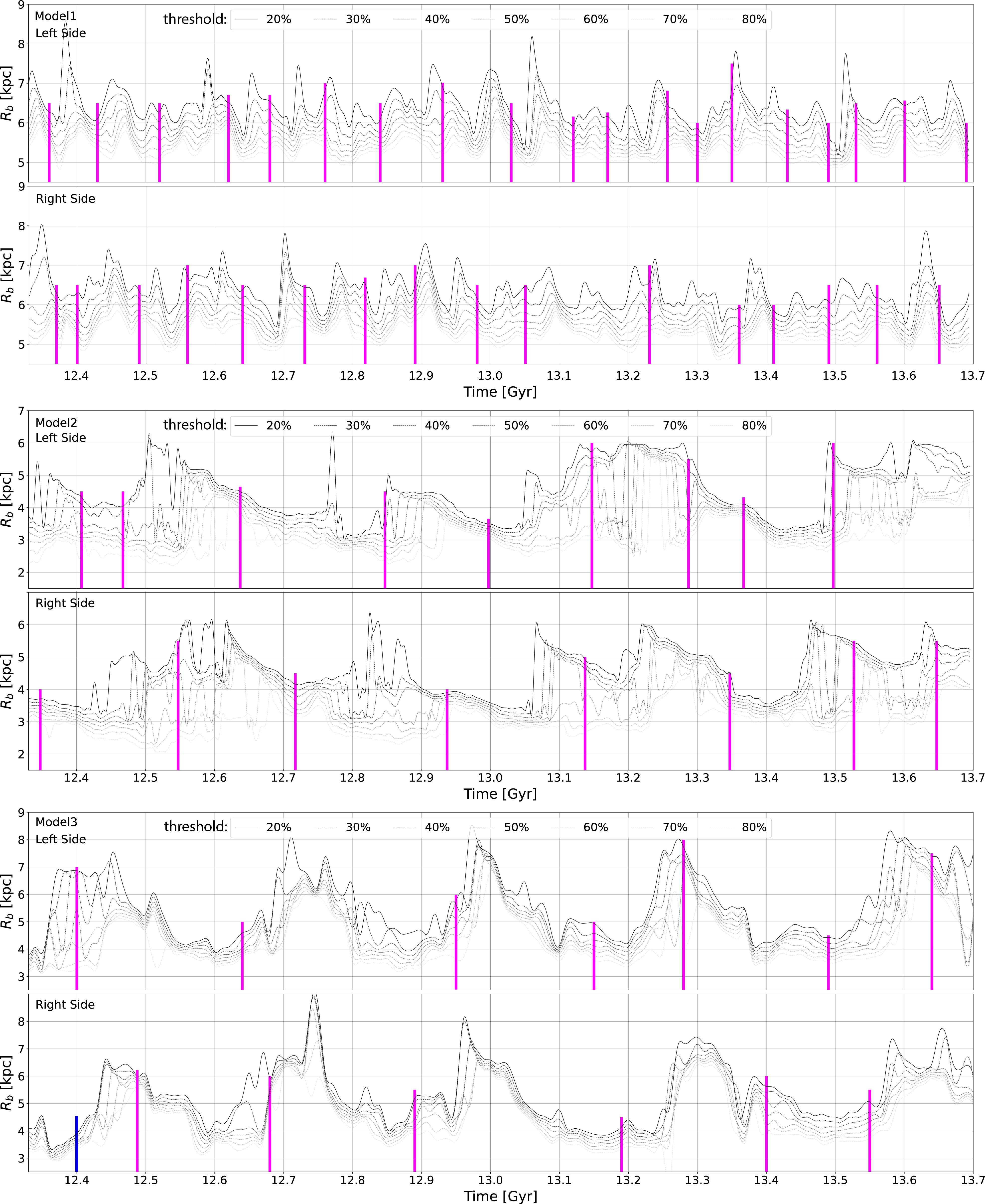}
\caption{
Variation in bar length with time, as in Fig.~\ref{fig:model3_length}. The three blocks of two rows show Model1 (top), Model2 (middle), and Model3 (bottom). Each block shows the Left Side (top) and Right Side (bottom) of the bar, as indicated. The vertical lines show the times when good matches to the \Gaia DR3 data are achieved, some of which were presented in Fig.~\ref{fig:matches_all}. The higher frequency of oscillations in the top panels reflects the fast bar of Model1, causing it to meet the spiral arms more often. From the number of wave packets seen in Model2 and Model3, it is clear that these bars are both slower than that of Model1, while the Model2 bar is the slowest.  
}
\label{fig:matches_t}      
\end{figure*}

\subsection{Matches to data, considering $v_R(x,y)$ inside solar circle}
\label{sec:loosly}

Focusing mostly on the $v_R$ structure inside the solar circle, we identified good matches to the \Gaia DR3 $v_R$ field, by visually inspecting all snapshots from out three simulations in the studied period of 1.37 Gyr. A representative sample of those is displayed in Fig.~\ref{fig:matches_all}, with the \Gaia DR3 data shown in the lower-leftmost panel. It is remarkable that Model1, with a bar size $R_b\approx5.2$~kpc, gives similar $v_R$ morphology as a bar as short as $\sim3$~kpc (Model2), or Model3's $\sim3.6$~kpc bar. This is, however, not surprising since we know that the fluctuation in bar parameters with time is strongly dependent on the strength of spiral structure outside the bar region \citep{hilmi20}. We scaled the distances in each model\footnote{In reality we did not have to scale Model1 and Model3, since they happened to match the $v_R(x,y)$ field straight out of the box.} so that we could obtain good matches to the observed $v_R(x,y)$ field. This has naturally resulted in an arrangement, such that smaller bars are accompanied by stronger spiral structure (Model2 and Model3), as reported in Sec.~\ref{sec:sp} {and seen in Fig.~\ref{fig:fourier} (see dominant m=2 mode)}. 

In Fig.~\ref{fig:matches_t} we show all the times at which good matches to the data are obtained for our three models (some of which were displayed in Fig.~\ref{fig:matches_all}), in order to understand if they always correspond to a certain bar-spiral orientation. The three blocks of two rows each, show the time evolution of the bar half-length for Model1 (top), Model2 (middle), and Model3 (bottom). The vertical lines indicate the times when good matches to the \Gaia DR3 data take place. The higher frequency of oscillations in the top panels reflects the fast bar of Model1, causing it to meet the spiral arms more often. From the number of wave packets seen in Model2 and Model3 it is clear that both of these bars are slower than that of Model1, while Model2's bar is the slowest. 

It can be seen from Fig.~\ref{fig:matches_t} that, in the same period of time, many more matches are found for Model1 than for the other models, as it should be expected if its bar encounters the spiral arms more often, as discussed above. For all models, the number of good matches is roughly equal to the number of peaks, i.e., to first order, it is expected that a good match occurs when the bar-spiral orientation is the same. But how do we determine the bar-spiral orientation? 

As discussed in Sec.~\ref{sec:modes}, the spiral seen in the stellar mass (Figures \ref{fig:xyt} and \ref{fig:match}) results from the constructive interference of all spiral modes of different multiplicities overlapping at a given time just outside the bar. We can see that in most matching time outputs from Model1 (20 out of 35) and Model2 (12 out of 17), the bar is to the right of the nearest peak, indicating it is in the process of separating from the spiral. 

However, only 5 out of the 14 matches from Model3 are for a bar disconnecting from a spiral, according to the above criterion. This includes our best match (see Fig.~\ref{fig:match}), shown by the blue vertical line in the bottom panel of Fig.~\ref{fig:matches_t}. Upon another inspection of the bottom panels of Fig.~\ref{fig:matches_all}, we can see that the Model3 matches (mostly focused on the upward positive $v_R$ leading arm, stemming from the positive $v_R$ lobe), have a common flaw outside the solar radius: a trailing positive $v_R$ arm extends also downward along the solar circle, which is not seen in the \Gaia DR3 data, nor in the other two models. This may be due to a spiral mode present in this simulation, that either does not exist in the MW, or is simply not as strong as the one causing the upward $v_R$ arm. In other words, the bar in Model3 is {\it connecting to one spiral mode while disconnecting from another}, as discussed in Sec.~\ref{sec:modes} and seen in the bottom left panel of Fig.~\ref{fig:match}. While this would also happen for the other two models, the difference is that these two modes in Model3 are of similar strength, judging from the similar response seen in the radial velocity field.

This expectation is confirmed in Fig.~\ref{fig:fourier}, which shows the Fourier components estimated from the face-on disc density for modes m=1-m=4 as functions of time. We can see that in the range 7.2-9.3 kpc, which is where the positive $v_R$ arm that stretches downward in most matching snapshots of Model3 (bottom row of Fig.~\ref{fig:matches_all}) is located, and which is not seen in the data, results from these odd modes. Indeed, Model3 shows the strongest m=1 and m=3 modes, corresponding to one-armed and three-armed spiral structure. The matches to the data appear to happen near m=1 and/or m=3 maxima, including the best match shown in Fig.~\ref{fig:match} (blue vertical). This suggests that the MW lacks such strong m=1 or m=3 modes in the radial range shown.

Although we found the best match to the data in Model3 considering the overall radial velocity field, Model1 and Model2 are consistently showing better matches to both the upper and lower right quadrants of the $v_R(x,y)$ plane simultaneously. It should be noted that the spiral structure responsible for the radial velocity outside the solar circle is expected to be due to yet slower moving modes, different from the ones reaching the bar, which further complicates the problem.

We conclude that most likely the MW near bar side is in the process of disconnecting from a spiral arm, even though our best match is for a connecting one according to the bar length fluctuation with time (Fig.~\ref{fig:matches_t}), but a disconnecting one according to the overdensity seen in the lower left panel of Fig.~\ref{fig:match}. Again, this complication is due to the presence of multiple modes and their interference as a function of time. More work is needed to understand better this behaviour. 

\section{Discussion and conclusions}

In this work we used three MW-like simulations of galactic discs to study the \Gaia DR3 radial velocity field, $v_R(x,y)$. For all models we examined the last 1.37 Gyr of evolution, using frequent time outputs, from 4.5 to 10 Myr depending on the simulation. This allowed to resolve well the $v_R(x,y)$ time variation caused by the interaction of multiple patterns in the disc, most importantly for this project - the bar-spiral periodic overlap. Our models' true bar lengths, resulting when the bar is separated from the inner spiral structure, are about 5.2, 3, and 3.6 kpc, for Model1, Model2, and Model3, respectively. 

Our results can be summarized as follows:

\begin{itemize}
\item We showed that the Galactic disc radial velocity field, $v_R(x,y)$, is a strong function of time, due to the relative orientation between the bar and spiral structure. The butterfly pattern in the bar region can thus vary dramatically both in shape and size over periods of a few tens of Myr (see Fig.~\ref{fig:xyt}).

\item Because of the above, the morphology of the \Gaia DR3 $v_R(x,y)$ field can be used to constrain the relative orientation between the bar and the inner spiral structure, although this is not straightforward. We found a very good match to the observations for a snapshot from our Model3, for a bar in the process of connecting to a spiral arm. However, identifying the times of all possible matches to the inner disc radial velocity field morphology, for all three models, we concluded that most likely the MW bar is in the process of disconnecting from a spiral, likely the Scutum-Centaurus (see discussion in Sec.~\ref{sec:loosly}).

\item The dominating factor distorting the bar's shape and decreasing its position angle with respect to the Sun-GC line, is the heliocentric distance uncertainty (Fig.~\ref{fig:angle}). While this affects the $v_R(x,y)$ morphology, the bar-spiral orientation produces more important variations in both the apparent bar length and the $v_R$ butterfly pattern (see Fig.~\ref{fig:xyt}).

\item We require a distance uncertainty of $\sigma_d<30\%$ in the models to match well the \Gaia DR3 data with $\sigma_d<20\%$, in order to reproduce the flatness of the transition between negative and positive $v_R$, which cannot be achieved with the 20\% error in the simulations (see Fig.~\ref{fig:gaia_comp}). This may suggest that the data distance uncertainties are underestimated {or that the bar angle is $20^\circ$, rather than the nominal value of $30^\circ$.}

\item We also considered the possibility that the bar angle is smaller than $30^\circ$, which would require smaller distance uncertainty in the simulation to align the transition in the bar butterfly pattern with the Sun-galactocentric line. We found that a bar at a 20$^\circ$ angle and $\sigma_d<20\%$ uncertainty can produce a good match to the data, although not as good as the 30$^\circ$ angle and $\sigma_d<30\%$ uncertainty (see Fig.~\ref{fig:gaia_comp}). More work is needed to explore the interplay between bar angle and distance uncertainties.

\item We showed that a range in bar lengths can reproduce the \Gaia DR3 radial velocity field (focusing on structure inside the solar circle; see Fig.~\ref{fig:matches_all}), provided smaller bars are accompanied by stronger spiral structure. Our simulations have bars with genuine lengths of about 5.2, 3.6, and 3 kpc and corresponding spiral structure overdensity of about 10\% for Model1 and 20-25\% for Model2 and Model3. Considering the MW spirals are expected to have $\sim20\%$ overdensity, it is tempting to conclude that the MW bar length is consistent with a bar of size below or around 4 kpc.

\item We calculated the Fourier components for our three models and found that Model3 has the strongest m=1 and m=3 spirals. This likely results in an additional feature in the $v_R(x,y)$ field (a positive $v_R$ arm stretching downward from the positive $v_R$ lobe), which does not exist in the data. This suggests that the MW lacks strong m=1 and/or m=3 modes.

Our conclusion that the MW bar's length is affected by an attached spiral is supported by the work by \cite{rezaei18}, which presented an extinction map using red clump and giant stars from the APOGEE survey, showing that the location of the Scutum–Centaurus spiral arm is likely connected to the bar’s near side (as shown in their fig. 4).
Another piece of evidence is the recent work of \cite{lucey23}, who measured the maximal extent of trapped bar orbits in APOGEE DR17 to extend to $\sim3.5$ kpc, very much consistent with our best-match model ($\sim3.6$ kpc).

One should be particularly careful in the interpretation of the velocity field when features are found along a line-of-sight from the solar position. It can be seen in the top-left panel of Fig.~\ref{fig:match} that the leading positive $v_R$ arm, stemming from the positive $v_R$ lobe associated with the bar, which our simulation so well reproduces with the 30\% distance error, is in fact broken at $(x,y)\approx(-2, 5)$ when no uncertainty is added. The 30\% distance error, however, causes the gap to disappear and match the data. A hint of this gap is found for distance uncertainty of $20\%$ in the top-left and middle-left panels of Fig.~\ref{fig:gaia_comp}, but we need probably less than 10-15\% error to identify it unambiguously, as in the bottom panels of the figure. Note that the 10\% distance uncertainty cut in the \Gaia DR3 data (bottom-rightmost panel) indeed seems to suggest the arm is broken at $(x,y)\approx(-4, 5)$ kpc.

While we found a very good match to the radial velocity field, there are other constraints that should be considered in future work. An obvious one is the tangential component of the velocity, $v_\phi(x,y)$, which would require to assume a rotation curve in order to subtract the Galactic disc rotation and exhibit the residuals. One can also consider the velocity dispersion, or velocity moments (e.g., \citealt{muhlbauer03, hey23}). 

The effect of the beat frequency between the bar and spiral structure needs to be explored, by keeping all other parameters the same. It is feasible that a lower beat frequency (i.e., a slow bar) would result in a stronger effect on the central $v_R(x,y)$ morphology, since fast bars will not have sufficient time for interaction with the spiral. This may be another reason why our Model1, which hosts a bar at the allowed lower limit in terms of the fraction of CR radius to bar length, $\mathcal{R}=R_{RC}/R_{bar}\approx1.02$ (to be compared to $\mathcal{R}\approx1.75$ for Model2's slow bar, see \citealt{hilmi20}), is not producing much variations in the size of the velocity field butterfly pattern, in addition to its weak spiral structure.

\end{itemize}

\section*{Acknowledgements}

We are grateful to the referee, Ronald Drimmel, for a very constructive and useful report. We also thank Robert Benjamin, Anthony Brown, Elena D'Onghia, Julio Navarro, and Jason Sanders for helpful discussions. We are thankful to the organizers of the Workshop "Mapping the Milky Way", hosted by the Lorentz Center in Feb 2023, which allowed for inspiring debates with specialists in the field, and where results from this work were first presented.
I.M. and B.R. acknowledge support by the Deutsche Forschungsgemeinschaft under the grant MI 2009/2-1. E.V. would like to express his sincere gratitude to the University of Sydney Physics Department for their welcoming faculty and invaluable support in research and academics.
T.B.'s contribution to this project was made possible by funding from the Carl-Zeiss-Stiftung.
This work has made use of data products from the European Space Agency (ESA) space mission \Gaia. \Gaia data are being processed by the \Gaia Data Processing and Analysis Consortium (DPAC). Funding for the DPAC is provided by national institutions, in particular, the institutions participating in the \Gaia MultiLateral Agreement. The \Gaia mission website is https://www.cosmos.esa.int/gaia. The \Gaia archive website is https://archives.esac.esa.int/gaia.




\bibliographystyle{mnras}
\bibliography{myreferences} 

\appendix

\section{Supplementary plots}

\begin{figure*}
\includegraphics[width=10.cm]{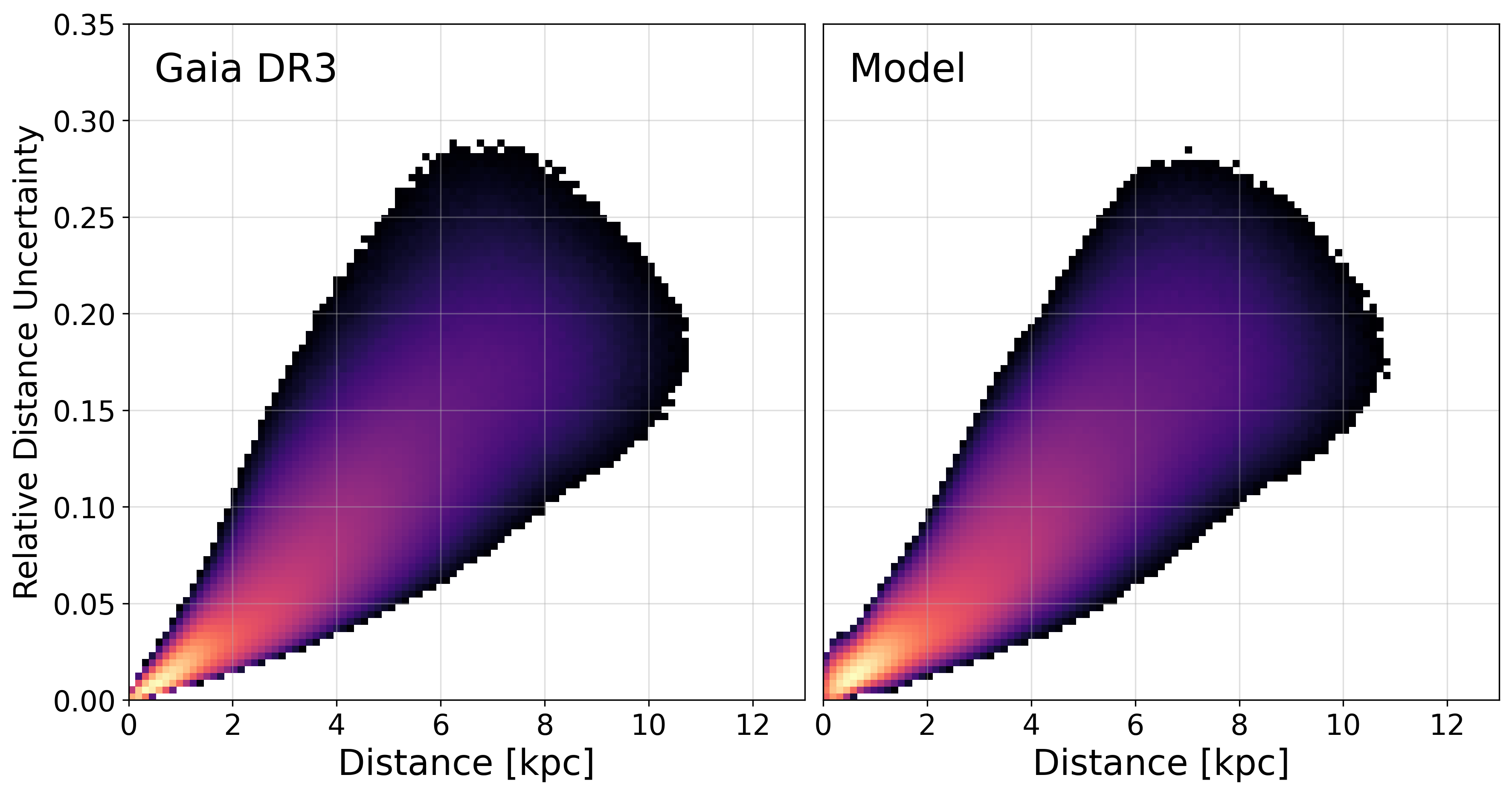}
\caption{{Left:} Distribution of relative heliocentric distance uncertainties ($\sigma_d\slash d$) versus distance, $d$, for the data. {Right:} Fit to the data on the left, using skewed Gaussian distributions for different distance bins. These are then interpolated to create a continuous probability density function dependent on distance as in the data. Yellow indicates higher density of stars.}
\label{fig:errMod}      
\end{figure*}

\begin{figure*}
\includegraphics[width=14cm]{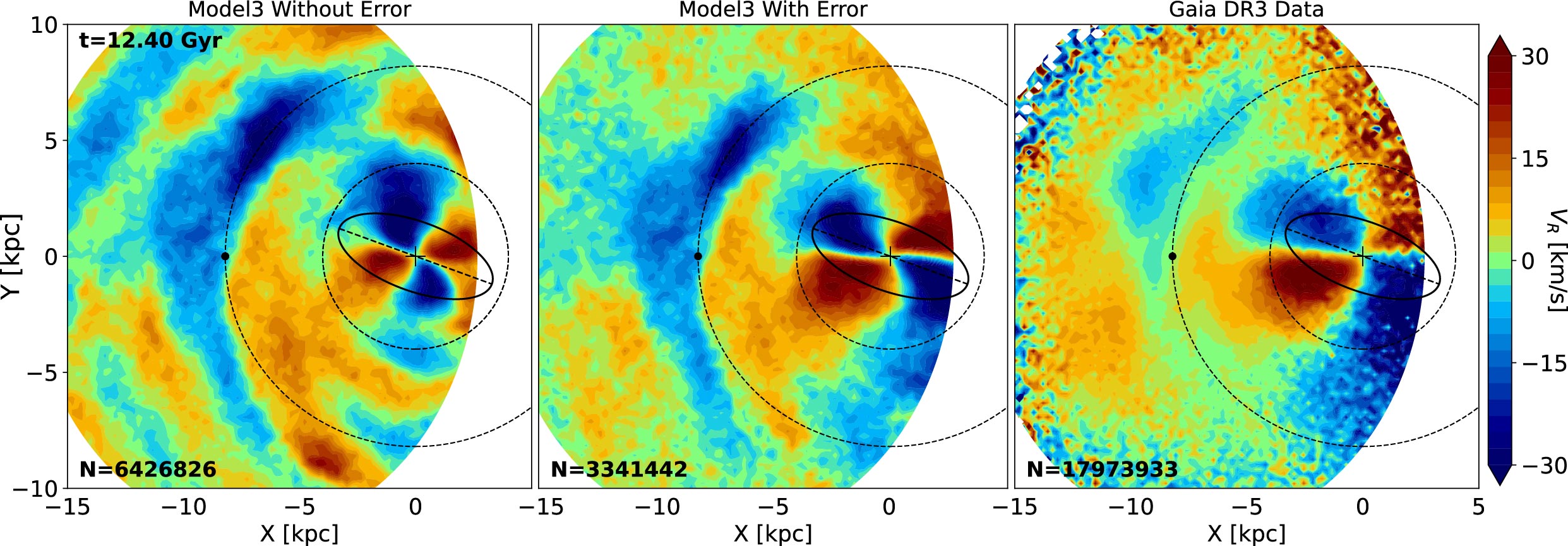}
\caption{
Match from Model3 to Gaia data, but using the Gaussian distance uncertainties as in the data, shown in Fig.~\ref{fig:errMod}. The data are still reproduced well, suggesting that the upward positive $v_R$ arm may have a break as in the model. In order to reproduce the flatness of the transition between the negative and positive $v_R$ lobes, a $20^\circ$ bar angle was required.
}
\label{fig:match2}      
\end{figure*}

\begin{figure*}
\includegraphics[width=17cm]{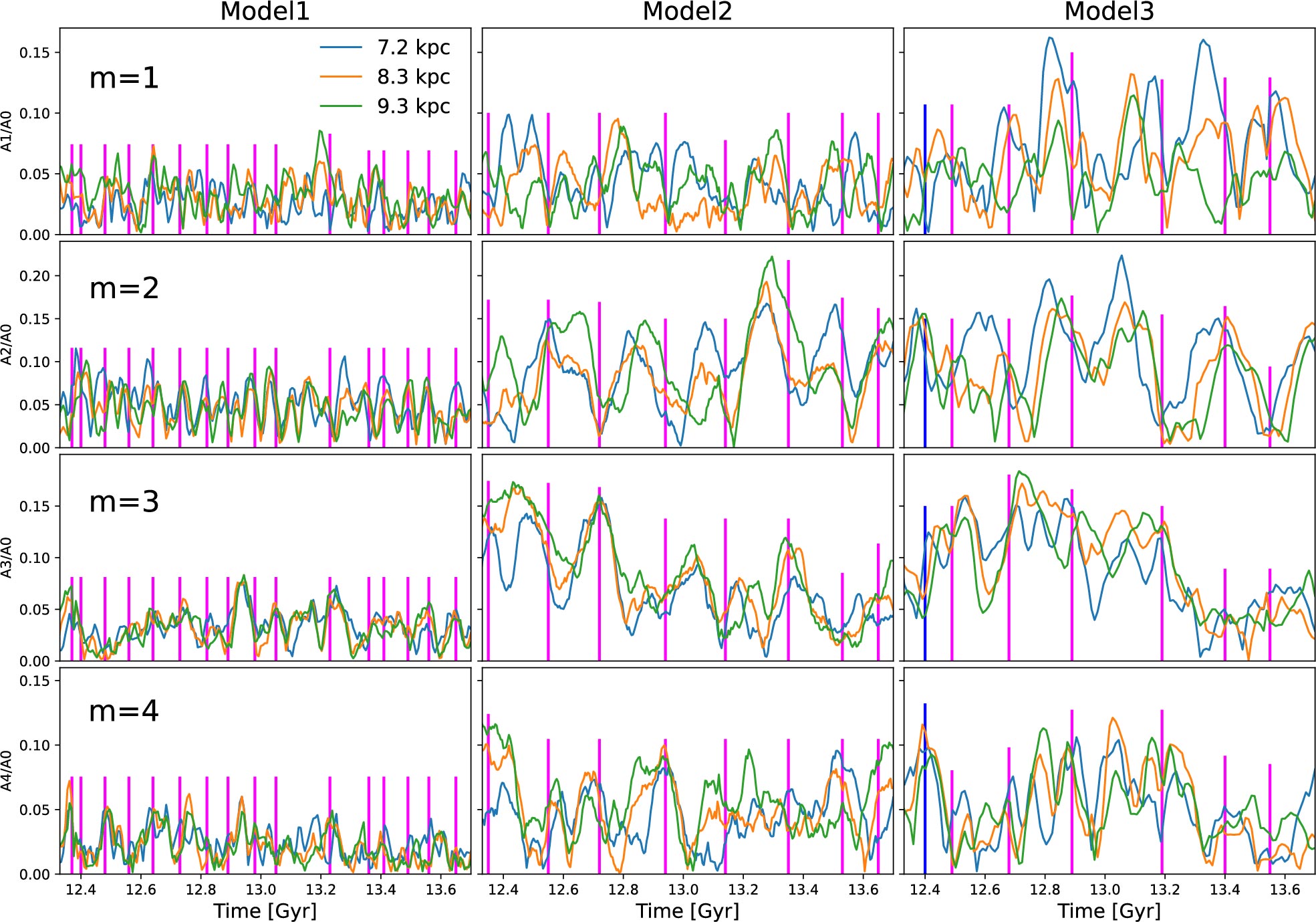}
\caption{ 
Fourier components estimated from the face-on disc density for modes m=1-m=4, as functions of time (see e.g., \citealt{athanassoula02}). Curves of different color show three radii in the range 7.2-9.3 kpc, as indicated. The vertical lines indicate the times when good matches to the \Gaia DR3 data are achieved for the Right side of the bar -- same as in the bottom panels of the three blocks found in Fig.~\ref{fig:matches_t}. Model3 shows the strongest m=1 and m=3 modes, corresponding to one-armed and three-armed spiral structure. The matches to the data appear to happen near m=1 and/or m=3 maxima, including the best match shown in Fig.~\ref{fig:match} (blue vertical). we can conclude that the positive $v_R$ arm, which stretches downwards in the matching snapshots of Model3 (bottom row of Fig.~\ref{fig:matches_all}), and which is not seen in the data, results from these odd modes. This would suggest that the MW lacks such strong m=1 or m=3 modes. 
}
\label{fig:fourier}      
\end{figure*}


\bsp	
\label{lastpage}
\end{document}